\documentclass{article}
\usepackage{arxiv}

\usepackage[utf8]{inputenc} 
\usepackage[T1]{fontenc}    
\usepackage{hyperref}       
\usepackage{url}            
\usepackage{booktabs}       
\usepackage{amsfonts}       
\usepackage{nicefrac}       
\usepackage{microtype}      
\usepackage{graphicx}
\usepackage{natbib}
\usepackage{doi}

\usepackage{multirow}
\usepackage{amsmath,amssymb}
\usepackage{amsthm}
\usepackage{mathrsfs}
\usepackage[title]{appendix}
\usepackage{xcolor}
\usepackage{textcomp}
\usepackage{manyfoot}
\usepackage{booktabs}
\usepackage{algorithm}
\usepackage{algorithmicx}
\usepackage{algpseudocode}
\usepackage{listings}
\usepackage{lscape}
\usepackage{threeparttable}
\usepackage{tabularx}
\usepackage{longtable}
\usepackage{threeparttablex}

\newcolumntype{L}{>{\raggedright\arraybackslash}X}

\title{Integrating behavior analysis with machine learning to predict online learning performance: A scientometric review and empirical study}

\date{March 28, 2024}	

\author{Jin Yuan \\
	Lanzhou University\\
		Lanzhou, China \\
	\texttt{yjin2023@lzu.edu.cn} \\	
	\And
	\href{https://0000-0002-5446-9758}{\includegraphics[scale=0.06]{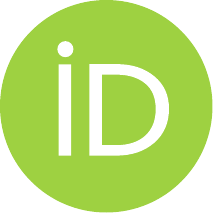}\hspace{1mm}Xuelan Qiu}\thanks{Corresponding author}\\
	Institute for Learning Sciences and Teacher Education\\
        Faculty of Education and Arts\\
	Australian Catholic University\\
	Brisbane, QLD, Australia, 4000\\
	\texttt{sherry.qiu@acu.edu.au} \\
	  \And
	  Jinran Wu\\
	Australian Catholic University \\
	  North Sydney, NSW, Australia, 2060\\
	\texttt{ryan.wu@acu.edu.au} \\
	  \And
	Jiesi Guo \\
	Australian Catholic University \\
	  North Sydney, NSW, Australia, 2060\\
	\texttt{jiesi.guo@acu.edu.au} \\
	  \And
	Weide Li \\
	Lanzhou University\\
		Lanzhou, China \\
	\texttt{weideli@lzu.edu.cn} \\
 	  \And
	You-Gan Wang \\
	The University of Queensland\\
		Brisbane, QLD, Australia, 4072\\
	\texttt{ygwanguq2012@gmail.com} \\
}

\renewcommand{\shorttitle}{Integration Framework to Predict Online Learning Performance}

\hypersetup{
pdftitle={Integration Framework to Predict Online Learning Performance},
pdfsubject={online.learning.prediction, computer.science},
pdfauthor={Jin Yuan, Xuelan Qiu, Jinran Wu, Jiesi Guo, Weide Li, You-Gan Wang},
pdfkeywords={Online learning, Performance prediction, Learning behavior analysis, Machine learning, Scientometric analysis},
}

\begin{document}
\maketitle

\begin{abstract}
The interest in predicting online learning performance using machine learning (ML) algorithms has been steadily increasing, partly due to the challenges posed by the COVID-19 pandemic. We first conducted a scientometric analysis to provide a systematic review of research in this area. The findings show that most existing studies apply the ML methods without considering learning behavior patterns, which may compromise the prediction accuracy and precision of the ML methods. This study proposes an integration framework that blends learning behavior analysis with ML algorithms to enhance the prediction accuracy of students' online learning performance. Specifically, the framework identifies distinct learning patterns among students by employing clustering analysis and implements various ML algorithms to predict performance within each pattern. For demonstration, the integration framework is applied to a real dataset from edX and distinguishes two learning patterns, as in, low autonomy students and motivated students. The results show that the framework yields nearly perfect prediction performance for autonomous students and satisfactory performance for motivated students. Additionally, this study compares the prediction performance of the integration framework to that of directly applying ML methods without learning behavior analysis using comprehensive evaluation metrics. The results consistently demonstrate the superiority of the integration framework over the direct approach, particularly when integrated with the best-performing XGBoosting method. Moreover, the framework significantly improves prediction accuracy for the motivated students and for the worst-performing random forest method. This study also evaluates the importance of various learning behaviors within each pattern using LightGBM with SHAP values. The implications of the integration framework and the results for online education practice and future research are discussed.
\end{abstract}

\keywords{Online learning \and Performance prediction \and Learning behavior analysis \and Machine learning \and Scientometric analysis}

\section{Introduction}
The rapid development of educational technologies in recent decades has dramatically changed the way we learn. Online learning, an essential form of distance education, enhances learners' experience by offering greater accessibility and flexibility than traditional classrooms. The COVID-19 pandemic has further brought attention to using advanced technologies for education to expand high-quality learning opportunities. Specifically, online learning enables large numbers of learners to access a broad range of courses, from computer science and business to humanities and sciences. For example, edX, one of the world's leading online learning platforms founded by Harvard and MIT, provides thousands of programs, and more than 81 million learners have enrolled on the platform by the end of 2023. These online courses are typically delivered through a variety of multimedia formats, including video lectures, interactive assignments, and collaborative forums, which are lower-cost and less constrained by locations and times. More importantly, online learning fosters a personalized learning approach by providing programs and courses that are tailored to learners' competencies, interests, and learning styles \citep{Klasnja2011}. This customization can promote learners' intrinsic motivations \citep{Alamri2020Using} and engagement \citep{Chiu2022}, leading to more effective learning outcomes. The transformative learning experiences enabled by technologies have contributed to the increasing awareness of making online learning an integral component of education. For example, the Activities to Support the Effective Use of Technology (Title IV-A) of the Every Student Succeeds Act provides funding to support schools in the United States, especially higher education institutions, to implement online learning programs. Title IV-A also sets out a statutory framework for the preparation and implementation of the National Educational Technology Plan \citep[NETP;][]{AEB}.

Since online learning is deemed to be crucial in shaping the future of education, there has been a compelling interest in exploring methods to predict students' online learning performances. The research interests are primarily driven by several important reasons. First, because of the inherent nature of online learning, the learning platforms save abundant educational data of students, ranging from basic information (e.g., age and gender) to learning activities throughout the learning journey (e.g., the number of hours devoted to learning courses, the total videos watched, and the frequency of forum posts). By leveraging predictive models, educators can identify patterns in students' academic behaviors and performance, which may indicate the necessity of early intervention for students at risk of academic challenges such as dropout \citep[e.g.,][]{Hu2014, Kuzilek2015, Mubarak2021deep}. As such, course completion and retention will be improved. Moreover, since the data can be conveniently collected over time, the information provides valuable insights for the improvement of online instruction. On the other hand, predicting students' performances in online learning can facilitate evidence-based policymaking on resource allocation and curriculum development for different educational stakeholders. For example, higher education institutes can optimize their strategies to ensure resources are allocated where they are needed most. In essence, predicting students' learning performances goes beyond foreseeing academic outcomes; it is a strategic approach to foster a more supportive and effective education system for learners.

In recent years, the applications of state-of-the-art machine learning (ML) in the field of education have surged because ML offers a powerful tool for predicting students' outcomes, such as dropout, retention, or withdrawal \citep{Hilbert2021}. These innovative methods have also been introduced to the research of online learning, where vast amounts of data would otherwise be challenging or even impossible to handle with traditional statistical methods. However, as is shown later in this paper, the majority of studies merely apply ML methods to the learning datasets without taking into account the characteristics of students' learning behaviors. Thus, the prediction accuracy and precision of ML methods are likely to be compromised. Moreover, the complex relationships between students' learning behaviors and their outcomes cannot be fully understood, potentially delaying the implementation of effective remedies.

The main aims of this research are twofold: First, to provide a systematic review of research about learning performance prediction using the ML methods and to identify research foci, emerging trends, and gaps, and second, to propose an innovative framework that integrates learning behavior analysis with ML methods and evaluates its effectiveness in improving prediction performance. To achieve these aims, this study conducts a scientometric review with the keyword co-occurrence analysis in the software CiteSpace \citep{Chen2006}. To the best of our knowledge, this paper is among the first to employ scientometric analysis in reviewing the topic. After introducing the integration framework, this study examines the performances of the proposed framework with a real online learning dataset retrieved from edX. Additionally, to showcase the advantages of the framework, this study also conducts an analysis employing ML methods directly on the empirical dataset and compares the results to those of the integration framework using comprehensive evaluation indices. This study contributes to the existing literature by offering a comprehensive review from a quantitative and visualizing perspective and shedding light on the evolution of research topics and trends. Moreover, the proposed framework enhances not only predictive performance but also the interpretability of students' learning behaviors, thereby contributing to more accurate and explanatory predictions in online educational settings. 

The structure of this paper is as follows: First, a scientometric review of predicting students' learning performances using the ML methods is presented. Second, the methods section describes the integration framework that blends learning behaviors analysis with ML algorithms is presented, followed by the brief introduction of the ML algorithms used in this study. Third, the empirical study describes the real data and analysis, and fourth, the results section presents the behavioral patterns in the real data and assesses the prediction accuracy and efficiency of the proposed framework. Furthermore, the integration framework was compared to the direct approach without behavioral analysis in terms of predictive performance. Finally, the implications of the results for online education practice and future research are discussed.

\section{A scientometric review of research on predicting learning performance using machine learning}
This section presents a scientometric review of studies on predicting learning performance using ML methods in the last 10 years.  To conduct the review, we chose the Web of Science Core Collection databases and applied an advanced topic search using a combined set of keywords of ``learning performance prediction'' or ``students-at-risk prediction,'' along with ``machine learning,'' and limited the search to articles published in English between 2014 and 2023. This search may include studies that focused on performance prediction using an ML approach in both online learning and traditional offline learning contexts. However, studies that predicted online learning without employing an ML method were excluded. A total of 299 publications (169 journal articles and 130 conference papers) were saved for further analysis.

This study utilized CiteSpace v.6.3.R1 advanced to conduct the scientometric analysis. The CiteSpace allows for quantitative analysis of bibliographic records by visualizing the knowledge maps and has been adopted in many systematic reviews of educational research topics \citep[e.g.,][]{Liu2022, Jing2020}. To identify the research foci and emerging trends, this study applied keyword co-occurrence analysis in CiteSpace to the publications collected from the previous search procedure. This analysis calculates the frequencies of keywords over the studied period and their co-occurrence relationship and generates a keyword co-occurrence network for predicting learning performance. Research foci are identified through high-frequency keywords in valid records while emerging trends are revealed based on high-frequency keywords that first appeared within the most recent years. As a result, the research gap is identified.

\subsection{Research focus analysis}
\begin{figure}
    \centering
    \includegraphics[width=1\linewidth]{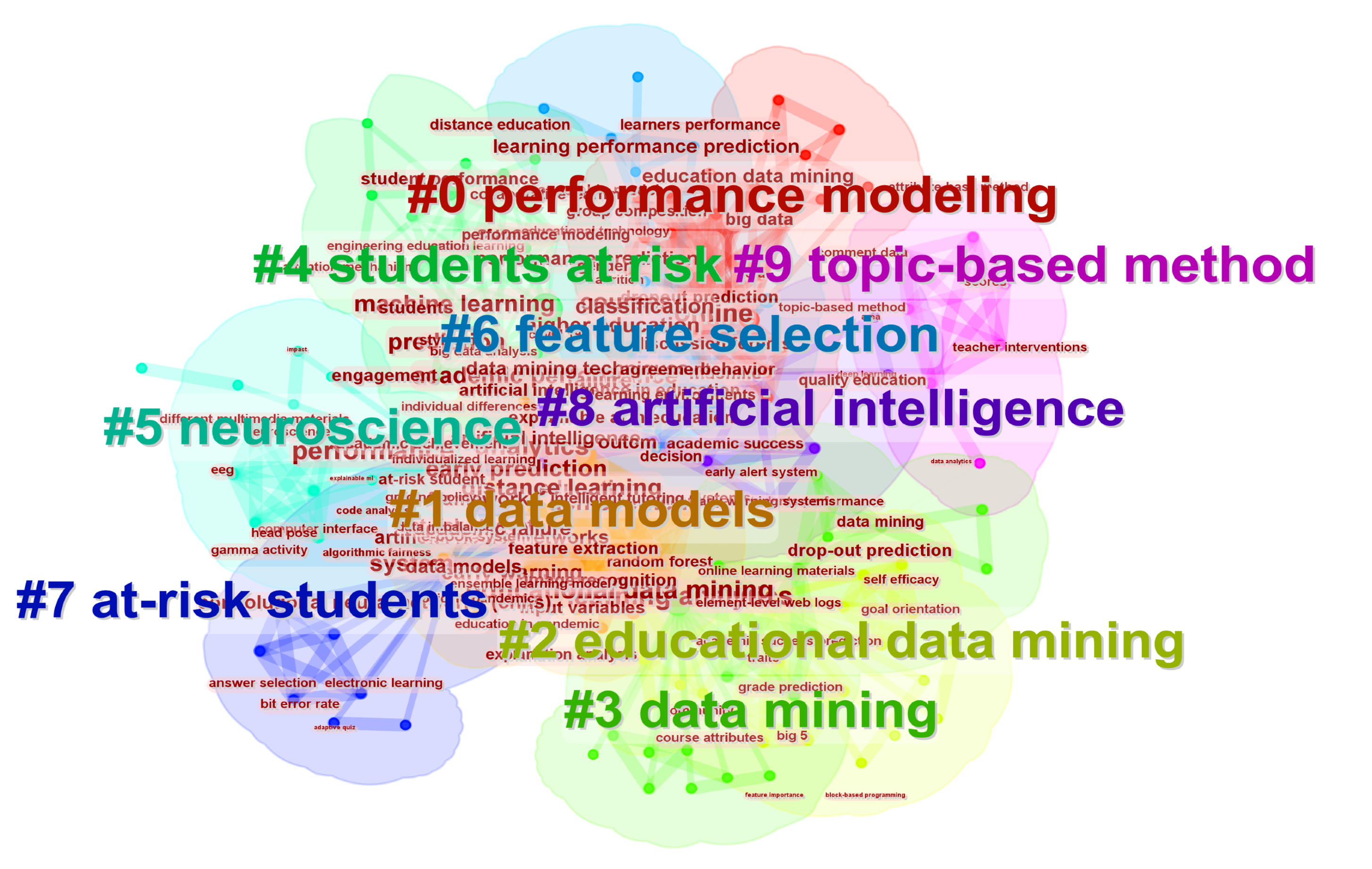}
    \caption{A visualization of keyword co-occurrence network using CiteSpace}
    \label{fig_keywords_cluster}
\end{figure}

Figure \ref{fig_keywords_cluster} presents the keyword co-occurrence network comprising 185 nodes and 739 links, from which ``performance modeling'' (with a total account of 86)  emerges as the most studied topic of learning prediction, including synonyms of ``performance prediction,'' ``achievement prediction,'' ``drop-out prediction,'' ``learning performance prediction,'' ``students at risk,'' and ``at-risk students.'' Other highly occurred keywords are  ``educational data mining'' (with a frequency of 60) which includes ``data mining,'' ``learning analytics,'' ``data mining algorithms,'' and ``classification,'' and "machine learning'' (with a frequency of 43). These high-frequency keywords indicate a research focus on utilizing educational data and advanced methods (e.g., learning analytics, data mining, and ML) to predict students' performance. For example, \cite{Gray2019} employs early engagement and ML to identify at-risk students who may fail. The study found that prediction accuracy can be as high as 97.2\%, demonstrating the effectiveness of ML algorithms in predicting student performance.

\subsection{Emerging trends analysis}
\begin{figure}
    \centering
    \includegraphics[width=1\linewidth]{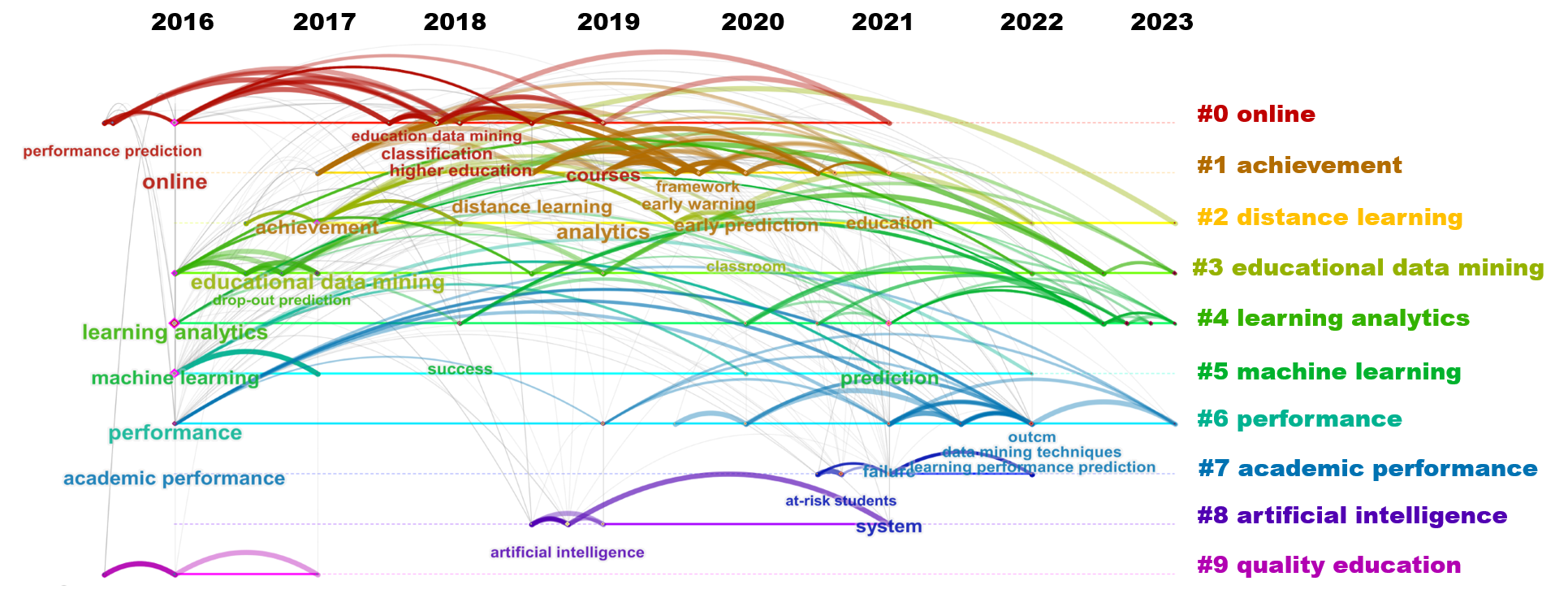}
    \caption{A timeline visualization of research focus (2014--2023)}
    \label{fig_keywords_timeline}
\end{figure}

Figure \ref{fig_keywords_timeline} presents a timeline visualization of the major clusters with highly cited works. Works conducted from 2014 to 2016 are not shown due to their relatively low citations. The hierarchy of clusters is primarily determined by their size, with labels decided by the keywords with the highest frequency in each cluster.

The largest cluster (\#0) focuses on online learning performance prediction, active from 2016 to 2021, followed by the second largest cluster of achievement (\#1), active from 2017 to 2021. Research in these threads primarily examines the use of education data mining to analyze data collected from Massive Open Online Courses (MOOCs) in higher education. Conversely, the third largest cluster of distance learning (\#2), which focuses on early warning and prediction, emerged in 2019 and remains active until 2023. This finding indicates a shift in online learning from MOOCs primarily offered by higher education institutions to distance learning platforms. Moreover, there has been a change in research focus from predicting performance in MOOCs to early identification of at-risk students in virtual learning environments. These changes are likely a response to the global COVID-19 pandemic, which forced a shift from traditional education to distance learning \citep{ALLILY2020}.

Moreover, educational data mining (\#3), learning analytics (\#4), machine learning (\#5), and performance (\#6) have spanned the entire study period. These results align with those depicted in Figure \ref{fig_keywords_cluster}. However, there is a shift in research focus from predicting success around 2018 to predicting failure, which began around 2021. Furthermore, the terms ``artificial intelligence (AI)'' and ``intelligent tutoring systems'' were first identified around 2019. It seems that the COVID-19 pandemic has increased students' difficulties and failures in learning, raising public concern. As a result, there has been a surge in research to develop alert systems and seek solutions using modern educational technologies such as distance learning and AI to tackle these challenges.

\subsection{Research gap}
Besides research focuses and emerging trends, the review also reveals the significant research gap in the field of predicting students' learning performances. In particular, among 299 studies, 244 studies were on learning in the traditional context in contrast to 55 on e-learning. Nevertheless, the keyword ``online'' was first detected in 2016 ``distance learning'' in 2019, and ``online learning'' in 2023, suggesting that online learning has a long history despite it becoming prominent in recent years because of the COVID-19 pandemic. Given the relatively small number of studies, there remains a need for further investigation into performance prediction in the context of online learning.

Although ``behavior analysis'' was not identified as a high-frequency term in CiteSpace, it has been implemented in several pivotal studies in the field, including the most highly cited work by \cite{AlShabandar2017}. To underscore this observation, we retrieved the top 20 most cited references identified in the scientometric analysis and conducted a survey based on objectives, data, and whether behavior analysis was involved. The results are shown in Table \ref{tab_review}, where 5 out of the 20 works include learning behavior analysis. It was found that research on the analysis of learning behaviors aims to identify key characteristics or patterns of learners from the learning behaviors features.  

For example, \cite{AlShabandar2017} conducted two experiments to assess the effectiveness of behavioral analysis combined with ML methods for predicting student outcomes in MOOCs. In the first experiment, all 12 features from the dataset, including 7 learning behaviors, were considered, while in the second, a subset of features was selected according to their rankings in importance. The results indicated that clickstream behavior ranked the highest and could effectively predict successful outcomes. This study demonstrates that using only a subset of behavioral features can yield satisfactory predictive results. Similarly, in the study by \cite{Amrieh2016}, 10 features, including 4 learning behaviors, were selected from a total of 16 features and utilized to predict students' academic performance. The study found that incorporating behavioral features led to over a 20\% improvement in accuracy compared to analyses without these features using traditional ML methods. Furthermore, when behavior features were integrated into ensemble methods, the improvement was even greater. Both \cite{AlShabandar2017} and \cite{Amrieh2016} applied feature selection techniques in behavior analysis.

In recent studies, clustering techniques have been employed to identify major patterns of behaviors. For example, \cite{qiu2022} applied the K-means clustering technique to classify 12 learning behaviors into three categories: learning preparation, knowledge acquisition, and consolidation behaviors. It was found that the performance of the method with behavior classification was superior to traditional classification methods.

While previous studies have demonstrated the effectiveness of behavior analysis in enhancing prediction accuracy \cite[e.g.,][]{Amrieh2016}, only 5 out of the 20 selected studies in Table \ref{tab_review} incorporated behavior analysis into their analyses, highlighting a significant gap in research exploring the integration of behavior analysis with ML algorithms. Hence, additional research is warranted to comprehensively examine the integrated framework. Such studies can offer deeper insights into students' learning behaviors and their influence on academic outcomes. Moreover, by enhancing prediction accuracy, the integrated framework has significant implications for improving educational practices in both traditional and online learning environments.

\begin{table}[ht]
\centering
\footnotesize
\caption{Review on the top 20 most cited references on predicting students' learning performances using machine learning methods}
\begin{threeparttable}
\label{tab_review}
\begin{tabularx}{\linewidth}{    
    >{\hsize=0.7\hsize}L  
    >{\hsize=1.5\hsize}L  
    >{\hsize=1\hsize}L  
    >{\hsize=0.8\hsize}L  
}
\hline
Article	&	Objective	&	Data	&	Behaviour Analysis	\\ \hline
\cite{AlShabandar2017}	&	To predict learning outcome in MOOCs	&	597,692 students with 7 behavior features and 5 demographic features	&	Yes (feature selection)	\\ \hline
\cite{Figueroa2020}	&	To predict dropout in an online statistics course	&	197 students with results of two assessments	&	Yes (dropout-prone and non-achievers)	\\ \hline
\cite{Abu_Zohair2019}	&	To predict performance in small dataset size	&	50 graduated students with 7 features	&	No	\\ \hline
\cite{Beaulac2019}	&	To predict university students’ academic success and major	&	38,842 students with 7 features	&	No	\\ \hline
\cite{Xing2016}	&	To predict dropouts in MOOCs using a temporal modeling approach	&	3,617 students	&	No	\\ \hline
\cite{Chui2020}	&	To predict at-risk students in a VLE	&	OULA dataset	&	No	\\ \hline
\cite{Akcapinar2019}	&	To develop early-warning system for at-risk students	&	76 university students with 7 types of learning processes data	&	No	\\ \hline
\cite{Adnan2021}	&	To predict course length taken by at-risk students	& OULA dataset	&	No	\\ \hline
\cite{Amrieh2016}	&	To predict student’s academic performance using ensemble methods	&	500 students with 16 features (4 demographic, 6 academic, 2 background, 4 behavioral)	& Yes (feature selection)	\\ \hline
\cite{You2016}	&	To identifying significant indicators using LMS data	&	530 college students, LMS data	&	No	\\ \hline
\cite{Mubarak2022}	&	To predict early dropout of VLE using interaction logs	&	OULA dataset	&	No	\\ \hline
\cite{Aggarwal2021}	&	To evaluate significance of non-academic parameters	& 6,807 students, academic and demographic data	&	No	\\ \hline
\cite{Al_Obeidat2018}	&	To analyze students’ performance using multi-criteria classification	&	1,044 students with 33 features	&	No	\\ \hline
\cite{Brinton2016}	&	To explore the relationships between video-watching behavior and quiz performance	& 6,450 students with 351,096 clickstreme events &	Yes (reflecting watching and revising watching)	\\ \hline
\cite{Hussain2019}	&	To predict academic difficulties	&	100 students, 6 digital exercises in each of 5 sessions	&	No	\\ \hline
\cite{Wakelam2020}	&	To predict performance in small cohorts	&	23 students, 3 features	&	No	\\ \hline
\cite{Huang2020}	&	To evaluate effects of classification methods and learning logs on prediction	&	7 datasets collected from 3 universities	&	No	\\\hline
\cite{Xing2019}	&	To predict dropout in MOOCs	&	3,617 undergraduate students and 12 features on learning behaviors	&	No	\\\hline
\cite{Waheed2020}	&	To predicte academic performance in VLE	&	OULA dataset	&	No	\\\hline
\cite{qiu2022}	&	To predict  performance in e-learning using learning process and behavior data	&	OULA dataset	&	Yes (learning preparation, knowledge acquisition, and consolidation behaviors)	\\ \hline
\end{tabularx}
\begin{tablenotes}
  \small
  \item \textit{Note}. MOOCs: Massive Open Online Courses; OULA dataset: Open University Learning Analytics dataset with 32,593 records and 32 features; VLE: Virtual learning environment. The references are listed in descending order of citations, with the most cited reference appearing at the top of the table.           
\end{tablenotes}
\end{threeparttable}
\end{table}

\section{Research Methods}
\subsection{The integration framework}
Unlike some studies \citep[e.g.,] []{Aggarwal2021, Amrieh2016} that investigated demographic characteristics (e.g., age, gender, and country) on academic performance, this work aims to predict students' performance through learning behavior analysis. The reason is that this work is anticipated to guide remedial solutions to enhance students' learning performance. However, variables such as age and gender are immutable. 
 
\begin{figure}
    \centering
    \includegraphics[width=0.8\linewidth]{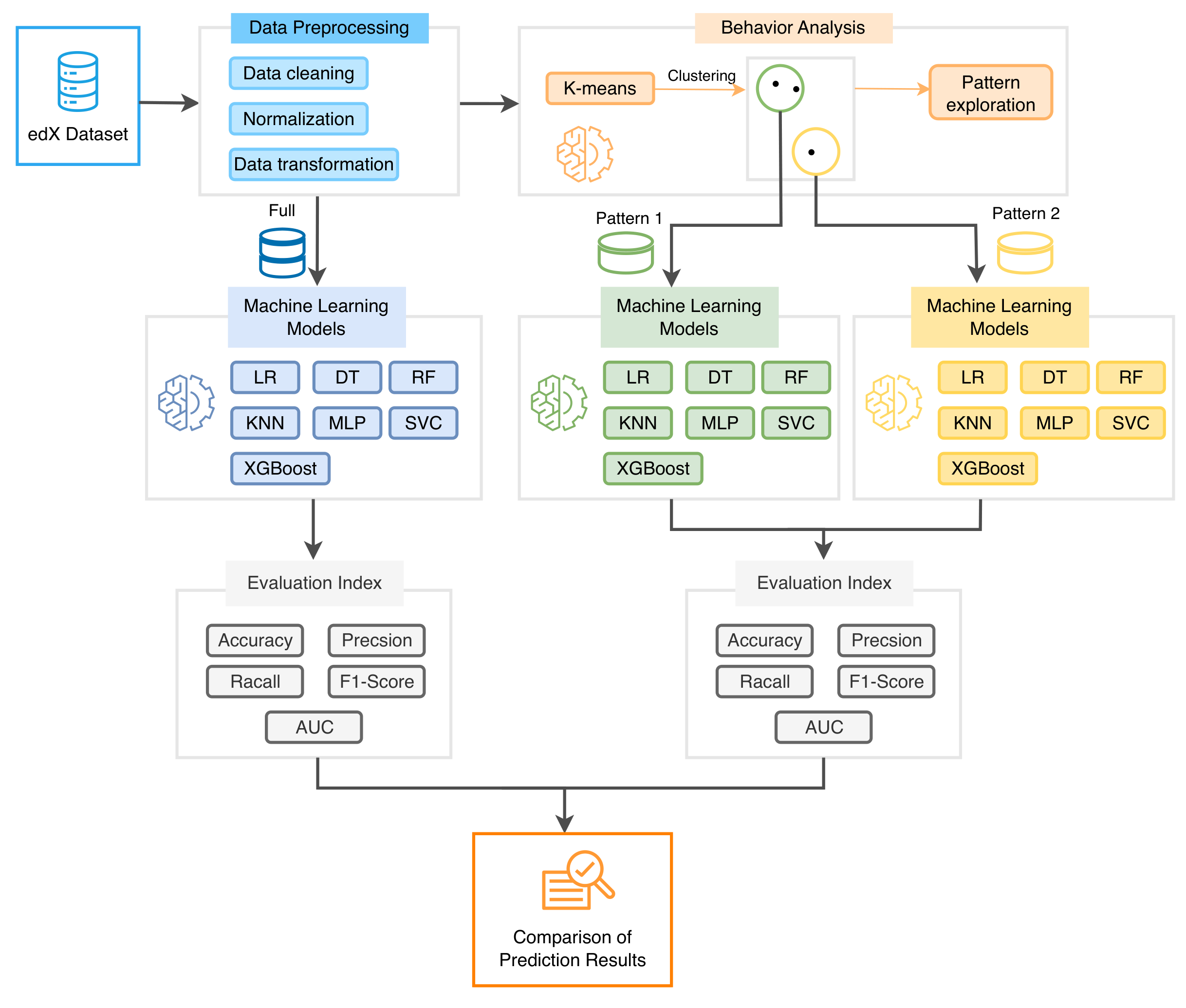}
    \caption{The integration framework for predicting learning performance (right side) against the direct approach without behavior analysis (left side)}
    \label{fig_framework}
\end{figure}

As depicted in Figure \ref{fig_framework}, the integration framework proposed in this study comprises two main stages: the first stage utilizes clustering analysis for online learning behaviors to explore learning categories or patterns, while the second stage employs various ML algorithms to predict students' performance within each identified category. In contrast to prior studies \citep[e.g.,][]{Cerezo2016}, which typically assess the significance of features and the associations between learning patterns and performances across the entire dataset, this framework examines these relationships separately for each identified category of students. The rationale behind this approach is that the likelihood of successful learning may significantly differ among students with different learning behavior patterns. Applying ML methods to each category is expected to enhance understanding of the relationship between learners' behaviors and their performance, thereby improving prediction accuracy.

The K-means algorithm, which partitions a dataset into a few distinct clusters based on similarity or distance to the clusters, is proposed for use in the first stage. The algorithm processes iteratively by assigning data points to clusters and updating the cluster centroids until convergence. For example, a well-known estimation method following the distance measure is silhouette analysis, which relies on the separation distance between the resulting clusters. Determining the optimal number of behavior patterns is a crucial yet challenging issue in the first stage. This study proposes to apply multiple indices to assess against a range of potential numbers of clusters.

In the second stage of the proposed framework, ML methods are applied to each identified learning pattern. This step is akin to the direct approach of implementing ML without behavior analysis, except that the latter utilizes the full dataset. Since the data is divided into different patterns of students, the issue of imbalanced classification may become more severe. Hence, this study proposes to employ advanced oversampling methods such as the Synthetic Minority Over-sampling Technique \citep[SMOTE;][]{Chawla2002SMOTESM} in this step.

\subsection{Machine learning algorithms}
This study implements varieties of ML algorithms to predict students' performances, including logistic regression (LR), decision tree (DT), random forest (RF), K-Nearest Neighbor (KNN), multilayer perceptron (MLP), Support Vector Classifier (SVC), and eXtreme Gradient Boosting (XGBoost). These algorithms are selected primarily for their popularity in various applications. It is worth noting that the performance of the algorithms can be significantly affected by the choice of parameters. Hence, it is crucial to carefully select and fine-tune hyperparameters. Additionally, one should be aware that every method has its drawbacks, aside from its strengths. Due to space constraints, this section briefly introduces the seven ML algorithms without details. Interested readers are referred to \citet{MLbookAlpaydin}. However, to help readers understand the algorithms, we provide conceptual illustrations in Figure \ref{fig_ML_conceptual_map} in the Appendix.

\subsubsection{The LR method}
The LR method was primarily developed for classification, both binary and multiclassification. Typically, it can be expressed in the form of the Sigmoid function. When the LR method is employed for prediction, students are predicted to be classified into the group of $Y=1$ (successfully receiving a certificate) if the successful probability is greater than 0.5, or into $Y=0$ (failing to receive a certificate) otherwise. 

\subsubsection{The DT method}
The DT method is a tree-like structure where internal nodes represent decisions based on specific feature values, branches represent decision outcomes and leaf nodes represent final predictions. It applies to both classification and prediction tasks and is popular for its interpretability. To construct and optimize decision trees, three selection measures—information gain, information gain ratio, and Gini index—are commonly used. In this study, we employ the Gini index, the most widely applied criterion for binary trees. The Gini index ranges from 0 (indicating perfect equality) to 1 (indicating maximal inequality), with a lower value indicating a better attribute for splitting the data and reducing inequality. The DT method, determines the optimal attribute for splitting data at each node, aiming to create more homogeneous subsets concerning the target variable. 

\subsubsection{The RF method}
The RF method constructs numerous DTs during training where each tree is constructed based on a randomly sampled subset of training data, and the final prediction is made by aggregating the predictions of all the trees. This ensemble approach enhances robustness and generalization performance, reducing the risk of overfitting by using a single DT. The RF method is deemed a popular ML method for its effectiveness in handling complex datasets and high-dimensional feature spaces. 

\subsubsection{The KNN method}
The KNN algorithm is a simple yet powerful method for prediction. In essence, the KNN algorithm calculates the distance between the new data point and all other points in the training dataset, typically using Euclidean distance in the feature space. The $K$ nearest neighbors are then determined based on these distances, and their class labels or values are used to predict the label or value of the new data point.

One of the key advantages of the KNN algorithm is its simplicity and ease of implementation. It does not require any assumptions about the underlying data distribution, making it particularly useful when dealing with complex or nonlinear relationships. Additionally, KNN can be effective in handling noisy data and is robust to outliers since it relies on local information from nearby data points.  

\subsubsection{The MLP method}
As a type of artificial neural network, the MLP consists of multiple layers of interconnected nodes, including an input layer, one or more hidden layers, and an output layer. Each node in the MLP is a neuron that processes information through weighted connections from the previous layer, applying an activation function to produce an output. Through a process called backpropagation, the MLP learns from data by adjusting the weights of connections iteratively to minimize the difference between predicted and actual outcomes.

MLP is popular mainly because of its ability to capture nonlinear relationships in data. With its multiple hidden layers, the MLP can learn complex patterns and relations from the input data. However, the performance of MLP models heavily depends on factors such as the choice of architecture (number of layers and neurons), activation functions, and optimization algorithms. Hence, careful tuning of these parameters is crucial to harness MLP for prediction. 

\subsubsection{The SVC method}
The idea of SVC is to find the optimal hyperplane that best separates different classes in the input data space. This hyperplane is positioned in such a way that it maximizes the margin between the nearest data points of different classes, thus enhancing the algorithm's generalization ability. SVC is particularly capable in handling high-dimensional data efficiently, making it suitable for applications with a large number of input features. Additionally, SVC is less susceptible to overfitting, especially in scenarios where the number of features exceeds the number of samples.

\subsubsection{The XGBoost method}
The XGBoost method is a powerful and efficient ML algorithm, particularly for prediction. It sequentially builds an ensemble of weak learners, typically decision trees, and optimizes them to minimize the overall prediction error. The method is popular due to several key features. Firstly, it incorporates regularization terms into its objective function, helping to reduce the overfitting problem and improve generalization. Secondly, it supports parallel processing, making it more efficient compared to traditional gradient boosting methods. Thirdly, it can handle missing values efficiently, which can simplify the data reprocessing.

\section{An empirical Study}
\subsection{Data and preprocessing}
The dataset for this empirical study was obtained from the edX platform \citep{HarvardX2014}, originally comprising 119,511 records for undergraduate students in 2014. As shown in Table \ref{tab_variables}, the dataset includes 11 variables, among which, there are three background variables (i.e., age, gender, and home country), seven behavior variables (e.g., whether access the courseware), and one performance variable (i.e., whether obtain the certificates). Before feeding it to the clustering and classification analysis, we preprocessed the dataset by removing missing values or identified problematic records, such as those flagged as ``incomplete'' or ``inconsistent'' \citep{HarvardX2014doc}. As a result, a total of 92,722 records from over 64 countries are used for subsequent analysis. 

\begin{figure}
\centering
\includegraphics[width=0.8\linewidth]{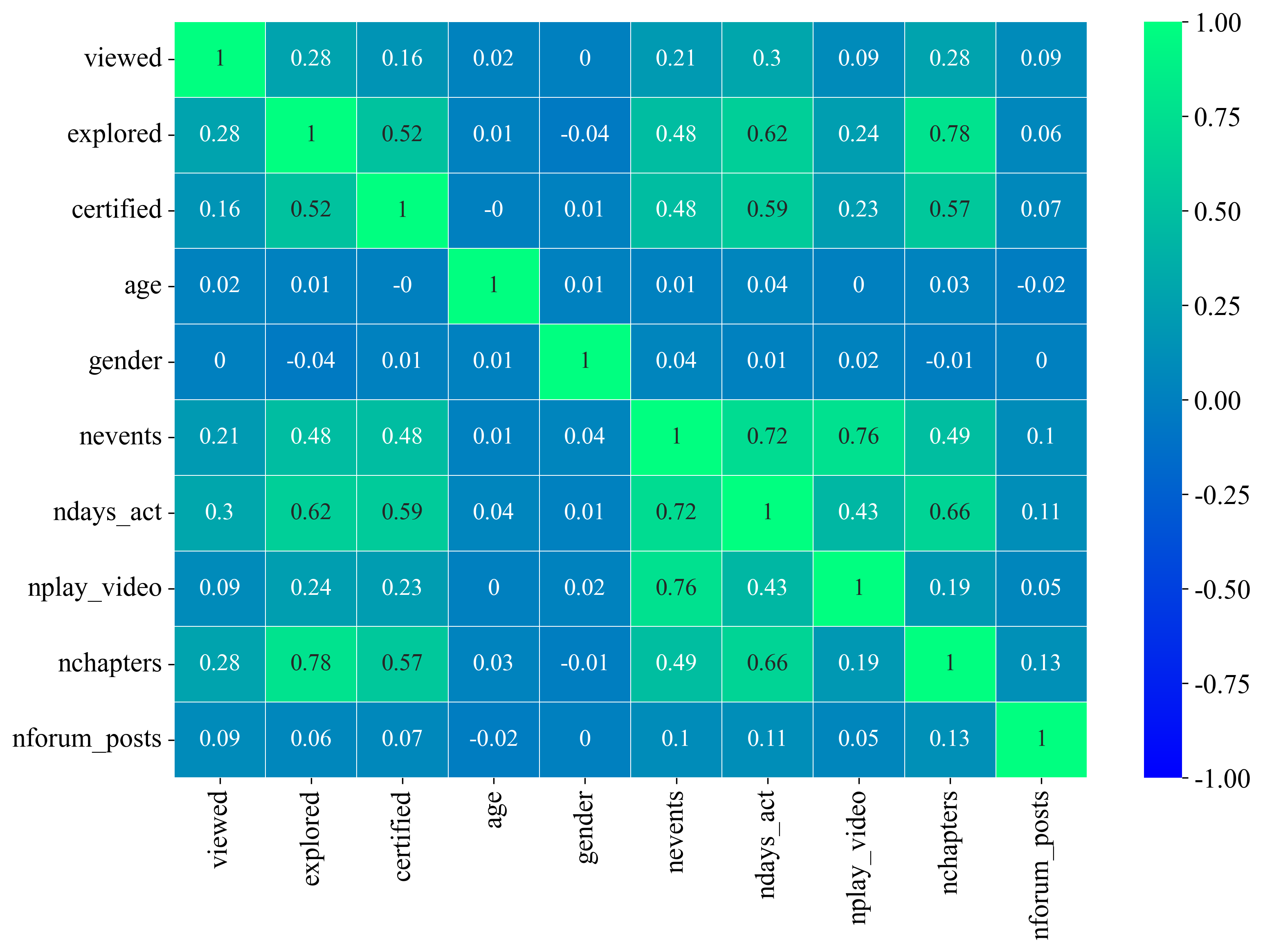}
\caption{Heat map for correlations between features}
\label{fig_heatmap}

\medskip
\begin{minipage}{0.8\textwidth}
{\footnotesize \textit{Note}: The heatmap does not include the ``country'' variable because it is a nominal variable with over 64 categories. Due to space constraints, only variable labels are shown, and descriptions can be found in Table \ref{tab_variables}.\par}
\end{minipage}
\end{figure}

We further standardized different features to the same scale using normalization and transformation techniques. Specifically, the behavioral variables were normalized, and the variables with string inputs were recoded into nominal variables. The descriptive analysis results are presented in Table \ref{tab_variables}, and the correlations between the variables are illustrated in the heat map shown in Figure \ref{fig_heatmap}. The sample has an average age of approximately 37 ($SD = 7.97$), with male students accounting for about 66.74\%. Weak correlations are found between gender and learning behaviors, as well as between age and learning behaviors. In general, the learning behaviors are moderately and positively correlated. The number of chapters studied (nchapter) and whether students accessed at least half of the chapters (explored) exhibit the highest positive correlation (.78), followed by the number of interactions (nevents) with the number of videos watched (nplay\_video) at .76, and with the number of days of interaction (ndays\_act) at .72.

\begin{table}[!ht]
\centering
\caption{Variables in the study}
\begin{threeparttable}
\label{tab_variables}
\begin{tabularx}{\linewidth}{
    >{\hsize=.3\hsize}X  
    >{\hsize=1.5\hsize}X  
    >{\hsize=1.2\hsize}X  
}
\hline
Type                        & Variable (Label)              & descriptive analysis      \\ \hline
\multirow{4}{*}{background} & age  (age)                    & $M = 37.75$, $SD = 7.97$ \\
                            & gender (gender)               & 66.74\% (Male), 33.26\% (Female) \\ 
                            & country (country)             &  $>64$ countries, the number of students ranges between 76 and 25,839  \\ \hline
\multirow{8}{*}{behaviors} & 1 whether accessed the courseware (viewed)    &  55.24\% (viewed) \\
                            & 2 whether accessed at least half of the chapters (explored) & 5.46\% (explored) \\
                            & 3 the number of days of interacting (ndays\_act)  & $M=3.66$, $SD = 6.83$ \\
                            & 4 the number of interactions (nevents)     & $M=151.21$, $SD=715.50$ \\
                            & 5 the number of videos watched (nplay\_video) & $M=32.88$, $SD = 166.75$  \\
                            & 6 the number of chapters studied (nchapters) & $M=3.00$, $SD=3.72$ \\
                             & 7 the number of forum posts (nforum\_posts)   & $M=0.01$, $SD=0.14$  \\  \hline         
performance                 & whether obtain the certificates (certified)    & 97.8\% (not certified)   \\ \hline
\end{tabularx}
\begin{tablenotes}
  \small
  \item \textit{Note}. \textit{M} = Mean, \textit{SD} = Standard deviation. 
\end{tablenotes}
\end{threeparttable}
\end{table}

\subsection{Analysis and evaluation metrics}
The key research questions in this study include:

1. What are the learning patterns of students in the real dataset from edX?

2. What are the performance metrics of the proposed integration framework in terms of prediction accuracy, precision, recall, F1 score, and other comprehensive evaluation indices? and

3. How does the predictive performance of the proposed integration framework compare to that of a direct approach that employs ML methods without behavior analysis?

Following the proposed integration framework, we first conducted a K-means clustering analysis to identify the learning patterns of students. The analysis considered three background variables and seven behavioral variables in Table \ref{tab_variables} and utilized the NbClust R package \citep{charrad2014nbclust} to explore clusters ranging from 2 to 8. The NbClust package offers 28 measures for determining the number of clusters built on various measures and clustering methods. In this work, the number of learning patterns was determined by assessing the frequency of explored clusters, with the optimal number of patterns identified based on the cluster that was most frequently selected by these measures.

Based on the identified learning patterns, we evaluated the performance of seven ML algorithms (i.e., LR, DT, RF, KNN, MLP, SVC, and XGBoost) on the datasets of students in each learning pattern after the training stage. Each model included the 10 variables for background and learning behaviors, as those in K-means cluster analysis. All these ML methods were trained with the Python scikit-learn library \citep{Pedregosa2011scikit} with a ratio of $7:3$ for training and testing datasets. 

The outcome variable examined in this study was ``certified,'' with students who were not certified constituting an overwhelming proportion of 97.8\%. This distribution indicates a substantial class imbalance. To address this issue, the SMOTE algorithm was employed in this study. Moreover, as mentioned earlier, the performance of ML methods may vary significantly due to the setting of hyperparameters. Due to space constraints, the hyperparameters used for each method in this study are shown in Table \ref{tab_hyperpar} in the Appendix.

To assess the prediction performance, four evaluation metrics are used in this study: Overall accuracy (accuracy), positive prediction accuracy (precision), true positive rate (TPR) or sensitivity (recall), and F1 score for balancing precision and recall. These metrics are defined as follows:
\begin{equation}\label{Eq:performance_metrics}
\begin{aligned}
\textnormal{accuracy} & =\frac{\textnormal{TP}+\textnormal{TN}}{\textnormal{TP}+\textnormal{FN}+\textnormal{FP}+\textnormal{TN}}, \\ 
\textnormal{precision} & =\frac{\textnormal{TP}}{ \textnormal{TP}+\textnormal{FP} }, \\ 
\textnormal{recall} & =\frac{ \textnormal{TP} }{ \textnormal{TP}+\textnormal{FN} }, \\ 
\textnormal{F1 score} & =\frac{ 2\times \textnormal{precision}\times \textnormal{recall}} {\textnormal{precision}+ \textnormal{recall} },
\end{aligned}
\end{equation}
respectively, where true positive (TP) represents positive samples predicted as positive by a given method, false positive (FP) represents negative samples predicted to be positive, false negative (FN) represents positive samples predicted to be negative, and true negative (TN) represents negative samples predicted as negative. 

In addition, the receiver operating characteristic (ROC) curve and the area under the curve (AUC) are used to evaluate the performance of the methods. The ROC curve assesses the trade-off between the TP rate and FP rate across various threshold settings, while the AUC quantifies the performance of a method by calculating the area under the ROC curve. In the ROC curve, a diagonal line represents random chance, while a line hugging the top-left corner indicates high sensitivity and a low FP rate. For AUC, values range from 0 to 1, with higher values indicating better model performance.
Moreover, violin plots, which combine the features of a box plot and a kernel density plot, are used to visualize the distribution of false positive rate (FPR) and true positive rate (TPR) for both low autonomy students and motivated students. If a method demonstrates high or nearly perfect prediction performance, the distribution of the FPR is expected to have median values near zero, and that of the TPR near 1. Additionally, the shapes of the distributions should be highly concentrated around their respective median values. 

One of the key research questions in this study is to compare the predictive performances of two approaches: the proposed integration framework, which blends ML methods with learning behavior analysis, and the direct approach without learning behavior analysis. As such, this study also applied the seven ML algorithms to the entire data without considering learning patterns. The hyperparameters for this analysis can be found in Table \ref{tab_hyperpar}, while other settings are the same as those of the integration framework. 

After selecting the best-performing algorithm to establish the online learning performance prediction model, the study proceeded to assess the relative importance of individual features within each identified learning pattern. While various methods exist for calculating feature importance, this study employed the LightGBM algorithm due to its renowned efficiency and speed in training. The seven behavioral features listed in Table \ref{tab_variables} were considered. In this study, we presented a ranked list of features based on their importance scores concerning each learning pattern. To enhance the interpretability of the models, this study further implemented the LightGBM algorithm using the Shapley Additive exPlanations (SHAP) method \citep{Lundberg2017} to measure each feature's contribution to the model prediction performance, both globally and locally. At a global level, the SHAP method analyzes the impact of each feature across the entire fitted dataset, providing a general understanding of feature importance, while on a local level, it reveals how each feature affects the model's performance for a specific data point. 

Finally, the dataset and the code for carrying out the analysis using the integration framework and the direct approach are freely available from the corresponding author upon request.

\section{ Results}
\subsection{Results of behavior analysis with the K-means clustering}
Due to space constraints, the results of K-means clustering analysis under different indices are shown in Table \ref{tab_res_clustering} in the Appendix. Excluding 4 graphical measures, it was found that 10 out of the 24 indicative measures in NbClust suggest an optimal number of clusters of two. To reveal the behavioral characteristics for these two identified clusters, Table \ref{tab_behavour_category} provides details regarding the sample size, learning outcomes (obtaining a certificate or not), and frequencies of the seven learning behaviors observed within each cluster. The students in cluster 2, although much smaller in number, exhibited significantly more intensive learning behaviors and achieved much better performance compared to their counterparts in cluster 1. 

Specifically, cluster 1 comprises 91,811 students, accounting for over 99\% of the total, where only 1,544 (1.68\%) students obtained a certificate. Students classified in this cluster displayed a relatively low level of interaction with the course material, with less than 50\% accessing the courseware and less than 6\% accessing half of the chapters. In contrast, among the 911 students in cluster 2, 53.24\% obtained certification upon course completion, a significantly higher rate than in cluster 1. All students in cluster 2 accessed the courseware, with over 90\% accessing at least half of the chapters. Moreover, these students engaged in, on average, approximately 13, 62, 44, 5, and 10 times more active days, interactions, videos watched, chapters studied, and forum posts, respectively, compared to those in cluster 1. To highlight the distinctions in their learning behaviors, the two identified clusters are henceforth referred to as ``low autonomy,'' and ``motivated,'' respectively, throughout the remainder of this paper.

\begin{table}[!htp]
\centering
\begin{threeparttable} 
	\caption{Comparisons of sample size, performance, and learning behaviors across clusters}
    \label{tab_behavour_category}	
\begin{tabular}{llllllllll}
\hline
        &              &                  & \multicolumn{7}{c}{Learning behaviors}                              \\ \cline{4-10} 
cluster & size(\%)  & certified(\%) & 1(\%)    & 2(\%)   & 3     & 4       & 5       & 6     & 7    \\ \hline
1       & 91811(99.02) & 1544(1.68)       & 42796(46.61) & 5039(5.49) & 3.28  & 94.61   & 23.06   & 2.89  & 0.01 \\
2       & 911(0.98)    & 485(53.24)           & 911(100)     & 846(92.86) & 41.72 & 5855.63 & 1022.94 & 14.25 & 0.10 \\ \hline
\end{tabular}
\begin{tablenotes}
    \small
    \item \textit{Note}. 1 = viewed; 2 = explored; 3 = ndays\_act; 4 = nevents; 5 = nplay\_video; 6 = nchapters; 7 = nforum\_posts. For learning behaviors 3 to 7, the values are their means of occurrence.  
\end{tablenotes}
\end{threeparttable}
\end{table}

\begin{figure}[!htp]
    \centering
    \includegraphics[width=1\linewidth]{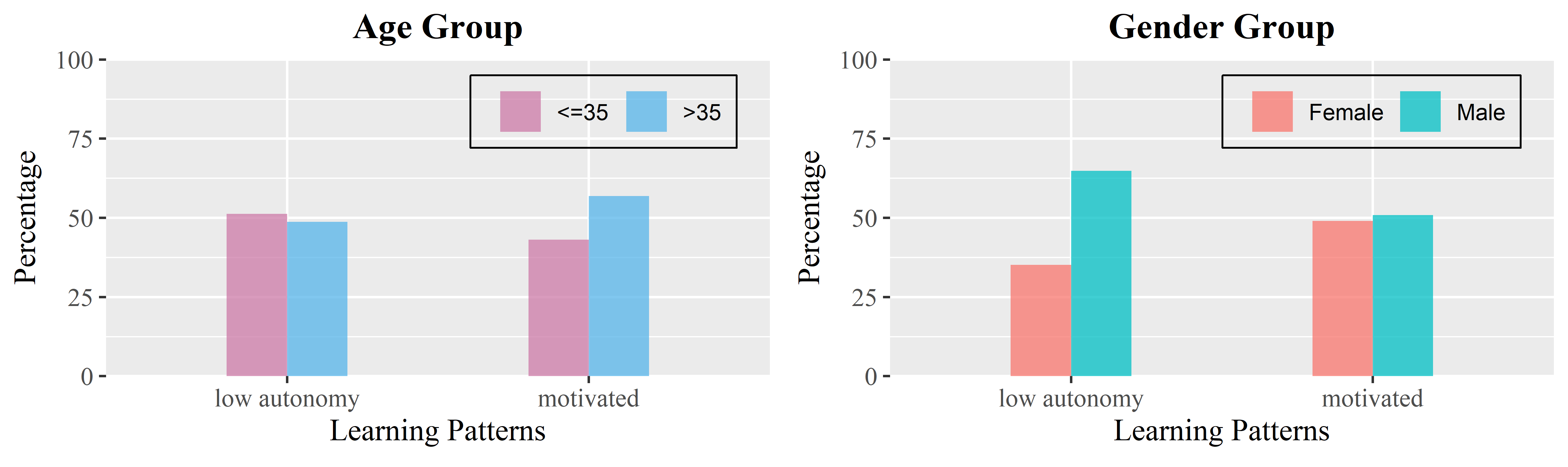}
    \caption{Comparison of learning patterns across age (left) and gender (right) groups}
    \label{fig_pattern_gender_age}
\end{figure}

It is interesting to explore potential age and gender differences in learning patterns. To this end, the age variable was first dichotomized using a cutoff of 35, which is near to the mean of the age variable. Figure \ref{fig_pattern_gender_age} shows the percentages of age groups (left panel) and gender groups (right panel) among low autonomy and motivated students. Specifically, among the low autonomy students, approximately 64.83\% are male and 51.25\% are younger than 35. Among motivated students, roughly 50.93\% are male and 56.86\% are aged older than 35. Subsequently, a Chi-square test of independence revealed significant differences between age groups ($\chi^2 (1) = 23.42$, $p = .000$) and gender group ($\chi^2 (1) = 75.66$, $p = .000$). Additionally, the effect sizes, measured by Cramer’s V, were found to be 0.016 for age and 0.029 for gender, respectively. Despite the statistically significant difference found in learning patterns among different age and gender groups, the effect sizes indicate a relatively small magnitude of difference.

\subsection{Results of online learning performance prediction using the integration framework}

To evaluate the prediction performances of the seven ML methods, Table \ref{tab_res_behavior_analysis} shows the accuracy, precision, recall, F1 Score, and AUC score for both low autonomy students (upper panel) and motivated students (lower panel). In general, the predictions for the low autonomy students by the ML methods are nearly perfect, and those for the motivated students are satisfactory. A closer inspection of the table indicates that the XGBoost method outperforms the other six ML algorithms. Specifically, for low autonomy students, the XGBoost method yields accuracy, recall, F1 score, and AUC score as high as 98.68\%, 99.10\%, 99.33\%, and 0.9929, respectively. For motivated students, XGBoost still maintains a higher level of performance than other methods, ranking the highest for accuracy (73.72\%) and AUC score (0.7960), while being the second highest for precision (73.33\%) and F1 score (70.97\%). In contrast, the other six methods, although showing promising prediction performance for low autonomy students, exhibit at least one performance index below 70\% when predicting outcomes for motivated students. In particular, the RF method has the lowest accuracy (91.64\%) for the prediction of low autonomy students while the DT method had the lowest recall rate (50.0\%) and F1 score (61.54\%), and AUC (0.6565) for the motivated students.

\begin{figure}[!ht]
    \centering
    \includegraphics[width=1.0\linewidth]{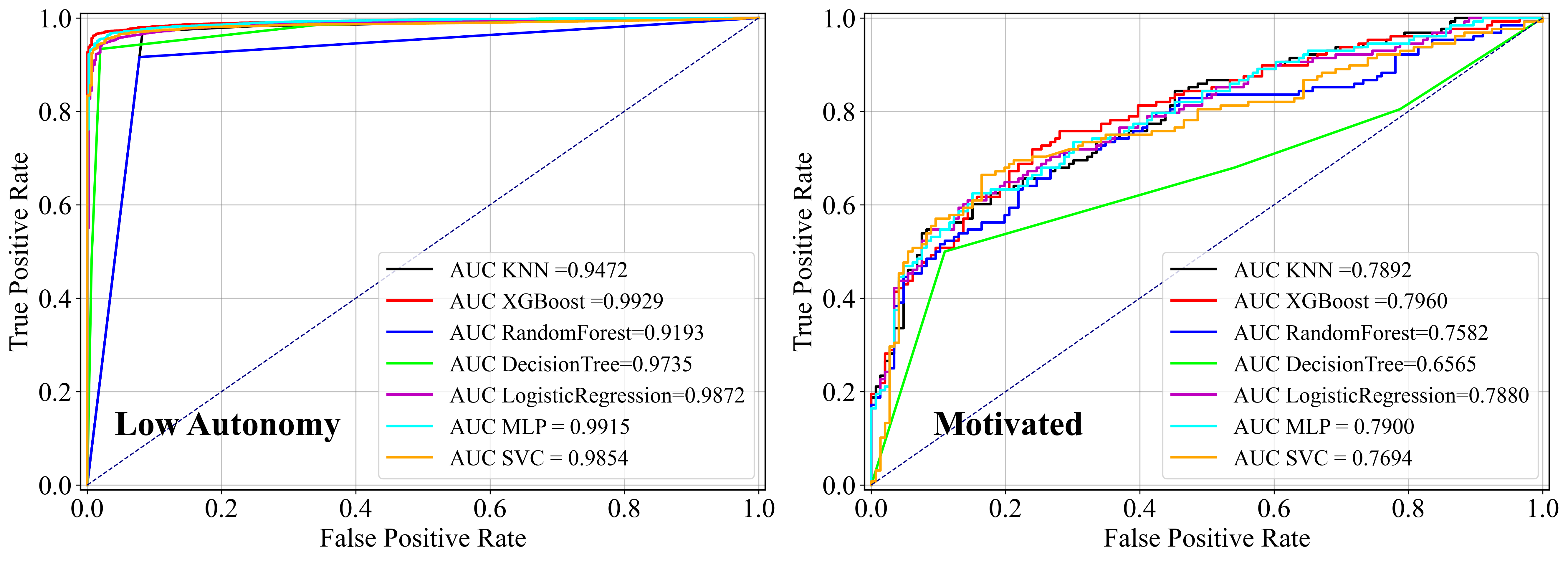}
    \caption{Receiver operating characteristic curves with area under the curve (AUC) scores of the machine learning algorithms for the low autonomy students (left) and the motivated students (right) under the integration framework}
    \label{fig_roc_behavior_analysis}
\end{figure}

\begin{figure}[!ht]
    \centering
    \includegraphics[width=1.0\linewidth]{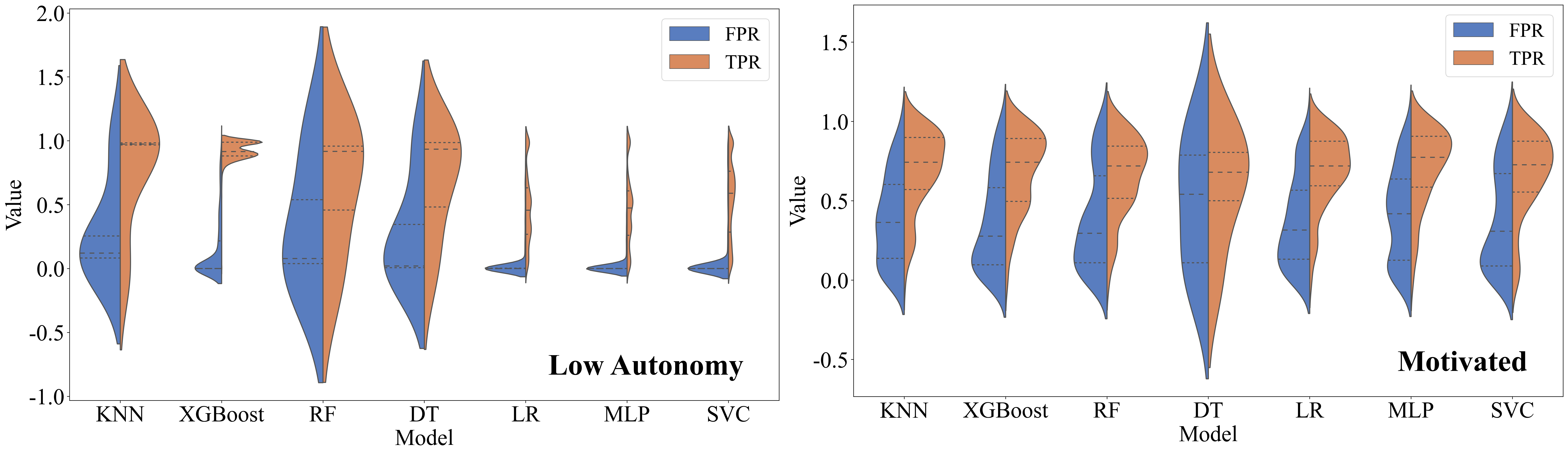}
    \caption{Violin plot shows the distribution of false positive and true positive rates for students who received certificates among the low autonomy students (left) and the motivated students (right) under the integration framework}
    \label{fig_violin_behavior_analysis}
\end{figure}

Additionally, Figure \ref{fig_roc_behavior_analysis} showcases the receiver operating characteristic (ROC) curves alongside the corresponding area under the curve (AUC) scores and Figure \ref{fig_violin_behavior_analysis} showcases the FPR and TPR distributions for the seven ML methods. For the low autonomy students on the left pane of Figure \ref{fig_roc_behavior_analysis}, XGBoost, LR, MLP, and SVC methods yield AUC scores exceeding 0.980, as evidenced by the clear distinction between FPR and TPR distributions under these methods on the left panel of Figure \ref{fig_violin_behavior_analysis}. Specifically, the FPR distributions for these four methods exhibit median values (indicated by the dashed line) and interquartile ranges (indicated by the two dotted lines) near zero, while TPR distributions are close to 1, indicating highly satisfactory prediction performance. Conversely, for motivated students on the right panel of Figure \ref{fig_roc_behavior_analysis}, the seven methods yield AUC scores ranging from 0.6565 (DT) to 0.7960 (XGBoost), which is confirmed by the blurred distinction between FPR and TPR where both the median values and interquartile ranges of FPR deviate from zero, while those of TPR away from 1 in the right panel of Figure \ref{fig_violin_behavior_analysis}.

\begin{table}[!htp]
\centering
\caption{Machine learning performances from predictive models for learning patterns under the integration framework}
\label{tab_res_behavior_analysis}	
\begin{tabular}{llccccc}
\hline
 Pattern& Method                 & Accuracy(\%) & Precision(\%) & Recall(\%) & F1 Score(\%) & AUC  \\ \hline
 &logistic regression    & 95.64& 99.92& 95.64& 97.73& 0.9872
\\
 &decision tree          & 93.44& 99.96& 93.36& 96.55& 0.9735
\\
low &random forest          & 91.64& 99.86& 91.63& 95.57& 0.9193
\\
 autonomy &K-Nearest neighbor     & 97.20& 99.79& 97.36& 98.56& 0.9472
\\
 &multilayer perceptron &96.45&98.79& 96.45&97.32&0.9915
\\
 &support vector classifier & 96.59&98.77&96.59& 97.40&0.9854 \\
&extreme gradient boosting   & 98.68& 99.55& 99.10& 99.33& 0.9929\\
\hline
 & logistic regression    
& 71.17& 68.15& 71.88& 69.96&0.7880\\
 & decision tree          
& 70.80& 80.00& 50.00& 61.54&0.6565\\
 & random forest          
& 70.80& 68.18& 70.31& 69.23&0.7582\\
motivated& K-Nearest neighbor     
& 72.26& 75.00 & 60.94& 67.24&0.7892\\
& multilayer perceptron 
& 70.44& 68.50& 67.97& 68.24&0.7900\\
& support vector classifier & 73.72& 72.95& 69.53& 71.20&0.7694\\
& extreme gradient boosting   
& 73.72& 73.33& 68.75& 70.97&0.7960\\
 \hline
\end{tabular}
\end{table}

In the integration framework, each learning pattern yields a set of performance metrics, resulting in a total of $K$ sets of learning performance metrics, where $K$ is the number of identified clusters in the clustering analysis for learning behavior ($K=2$ in this case). Conversely, the direct approach generates only one set of performance metrics. To facilitate comparisons between the two approaches, we computed overall accuracy, precision, recall, and F1 score across $K$ clusters based on the predictions of class labels obtained from the integration framework, using their respective definitions as outlined in Equation \ref{Eq:performance_metrics}. These results are presented in Table \ref{tab_pre_merge01}. Subsequently, we compare these findings with those derived from the direct approach in the next section.

\begin{table}[!htp]
\centering
\caption{
Overall performances across learning patterns for machine learning algorithms based on the results of the integration framework}
\label{tab_pre_merge01}	
\begin{tabular}{lccccc}
\hline
Method                 & Accuracy (\%) & Precision (\%) & Recall (\%) & F1 Score (\%) & AUC \\ \hline
logistic regression    & 95.53& 99.76& 95.66& 97.67& 0.9816\\
decision tree          & 93.32& 99.90& 93.25& 96.47& 0.9633\\
random forest          & 90.20& 99.75& 90.21& 94.74& 0.9651\\
K-Nearest neighbor     & 97.89& 99.90& 97.94& 98.91& 0.9865\\
multilayer perceptron &95.94&99.81& 96.04&97.88&0.9853\\
support vector classifier & 96.53&99.85&96.59& 98.19&0.9842\\
extreme gradient boosting   & 99.18& 99.77& 99.40& 99.58& 0.9971\\
\hline
\end{tabular}
\end{table}

\subsection{Comparisons of performance prediction between the approaches with and without learning behavior analysis}

\begin{figure}[!htp]
    \centering
    \includegraphics[width=0.55\linewidth]{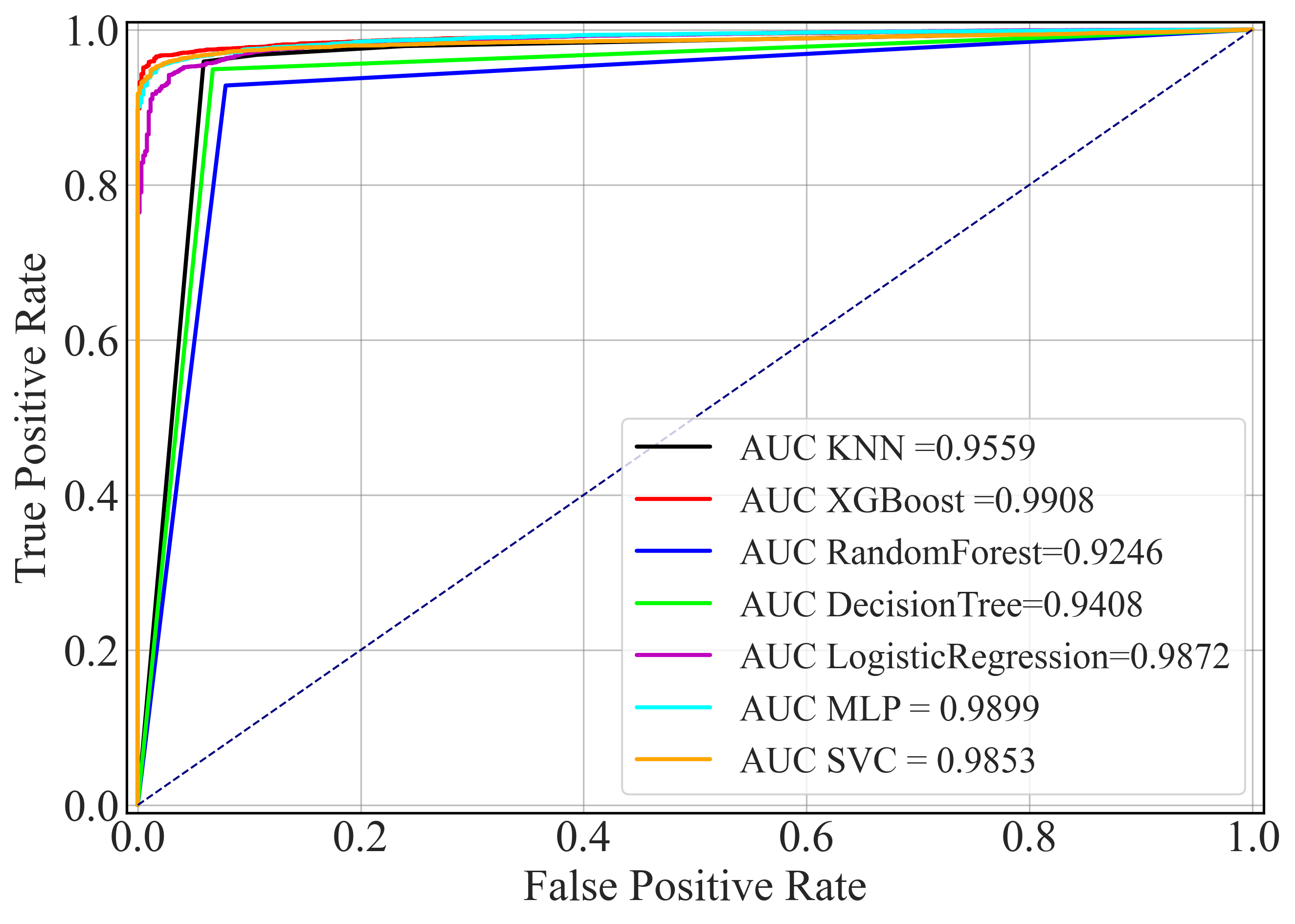}
    \caption{Receiver operating characteristic curves with area under the curve (AUC) scores of the machine learning algorithms under the direct approach without behavior analysis}
    \label{fig_roc_no_behavior_analysis}
\end{figure}

\begin{figure}[!htp]
    \centering
    \includegraphics[width=0.55\linewidth]{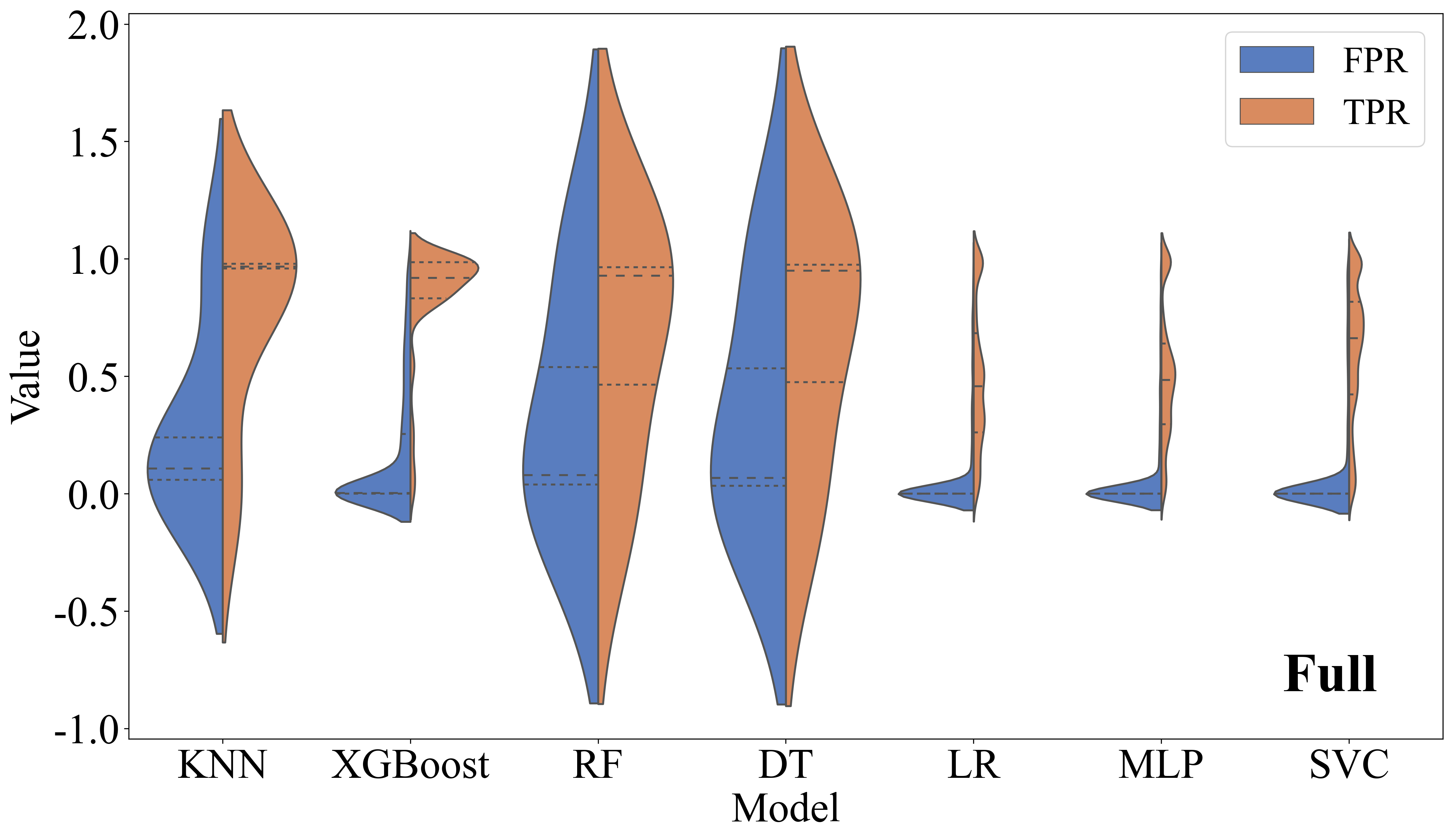}
    \caption{Violin plot shows the distribution of false positive and true positive rates for students who received certificates under the direct approach}
    \label{fig_violin_no_behavior_analysis}
\end{figure}

To showcase the advantages of the proposed integration framework, which incorporates learning behavior analysis, over the direct approach without such analysis, this section provides the results obtained through the direct approach and compares them between the two approaches. As illustrated in Table \ref{tab_res_no_behavior_analysis}, the accuracy rates range from 92.78\% to 97.58\%, precision rates from 98.11\% to 98.62\%, recall rates from 92.78\% to 97.58\%, F1 scores from 94.85\% to 97.94\%, and AUC scores from 0.9246 to 0.9908. The ROC curves for the seven ML methods under this approach are provided in Figure \ref{fig_roc_no_behavior_analysis}, and the FPR and TPR distributions are shown in Figure \ref{fig_violin_no_behavior_analysis}. Again, XGBoost performs the best, while RF performs the worst among the ML methods.

These results are compared to those from the integration framework, as shown in Table \ref{tab_pre_merge01}, where there is a range of values between 90.20\% and 99.18\% for accuracy, 99.75\% and 99.90\% for precision, 90.21\%, and 99.40\% for recall, 94.74\% and 99.58\% for F1 score, and 0.9633 and 0.9971 for AUC. The comparisons underscore that the XGBoost method, being the best-performing method, exhibits much better performance when learning behaviors are taken into account. Specifically, the accuracy rates for the integration framework are 99.18\% compared to 97.58\% for the direct approach, 99.77\% versus 98.62\% for precision, 99.40\% versus 97.58\% for recall rates, 99.58\% versus 97.94\% for F1 score, and 0.9971 versus 0.9908 for AUC score. Hence, the integration framework consistently demonstrates superior performance compared to the direct approach, especially when the learning behavior analysis integrates with the best-performing method (i.e., XGBoost).

\begin{table}[!htp]
\centering
\caption{Machine learning performances from the direct approach without behavior analysis}
\label{tab_res_no_behavior_analysis}	
\begin{tabular}{lccccc}
\hline
Method                 & Accuracy(\%) & Precision(\%) & Recall(\%) & F1 Score(\%) & AUC  \\ \hline
logistic regression    & 95.60& 98.38    & 95.60& 96.62& 0.9872\\
decision tree          & 94.86& 98.29   & 94.86& 96.14& 0.9408\\
random forest          & 92.78   & 98.11    & 92.78 & 94.85   & 0.9246 \\
K-Nearest neighbor     & 96.59& 98.40& 96.59& 97.25& 0.9559\\
multilayer perceptron &95.88&98.49& 95.88&96.82&0.9899\\
support vector classifier & 95.89&98.50&95.89& 96.83&0.9853\\
extreme gradient boosting   & 97.58& 98.62& 97.58& 97.94& 0.9908\\
\hline
\end{tabular}
\end{table}

We delve deeper to identify the learning pattern for which the integration framework exhibits greater improvement. To do so, we separated the whole sample into two groups utilizing the learning pattern labels for each student obtained from the integration framework and then computed the performance metrics with the predictions obtained from the direct approach for each group, respectively. 
The results are presented in Table \ref{tab_res_pattern_no_behavior_analysis} and compared with those in Table \ref{tab_res_behavior_analysis} from the integration framework.

Using the results from the direct approach as a baseline, the integrated framework enhances predictions by an average of 0.04\%, with a range of -3.78\% to 5.07\% for low autonomy students. In the case of predictions for motivated students, the average improvement is 12.64\%, ranging from -17.33\% to 84.71\%. In particular, the integration framework yields significant improvement for the RF method, the worst-performing method, by an average of 38.06\%. Therefore, the integration framework demonstrates its significance in improving prediction performance, especially for motivated students and with the worst-performing method (i.e., RF).

\begin{table}[!ht]
\centering
\caption{Separated performance for different learning patterns for machine learning algorithms  based on the results of the direct approach without behavior analysis}
\label{tab_res_pattern_no_behavior_analysis}	
\begin{tabular}{llccccc}
\hline
Pattern &Method                 & Accuracy (\%) & Precision (\%) & Recall (\%) & F1 Score (\%) & AUC \\ \hline
 &logistic regression    & 96.13& 98.67& 96.13& 97.09& 0.9866
\\
 &decision tree          & 95.38& 98.57& 95.38& 96.62& 0.9265\\
low &random forest          & 93.30& 98.44& 93.30& 95.38& 0.9140\\
autonomy &K-Nearest neighbor     & 97.67& 98.95& 97.67& 98.12& 0.9844\\
&multilayer perceptron &96.62&98.78& 96.62&97.42&0.9903\\
&support vector classifier & 96.35&98.80&96.35& 97.25&0.9869\\
&extreme gradient boosting   & 98.30& 98.97& 98.30& 98.54& 0.9936
\\ 
\hline
 & logistic regression    
& 55.76& 72.39& 55.76& 42.73&0.7600\\
 & decision tree          
& 60.48& 77.32& 60.48& 51.37&0.5775\\
 & random forest          
& 53.46& 75.17& 53.46& 37.48&0.5023\\
motivated & K-Nearest neighbor     
& 69.59& 77.85& 69.59& 66.26&0.9014\\
& multilayer perceptron 
& 59.93& 75.07& 59.93& 50.76&0.7105\\
& support vector classifier & 62.79& 76.63& 62.79& 55.61&0.7361\\
& extreme gradient boosting   
& 71.90& 76.11& 71.90& 70.10&0.7869\\
 \hline
\end{tabular}
\end{table}

\subsection{Results of feature importance for students with different learning patterns}
\begin{figure}
    \centering
    \includegraphics[width=1.0\linewidth]{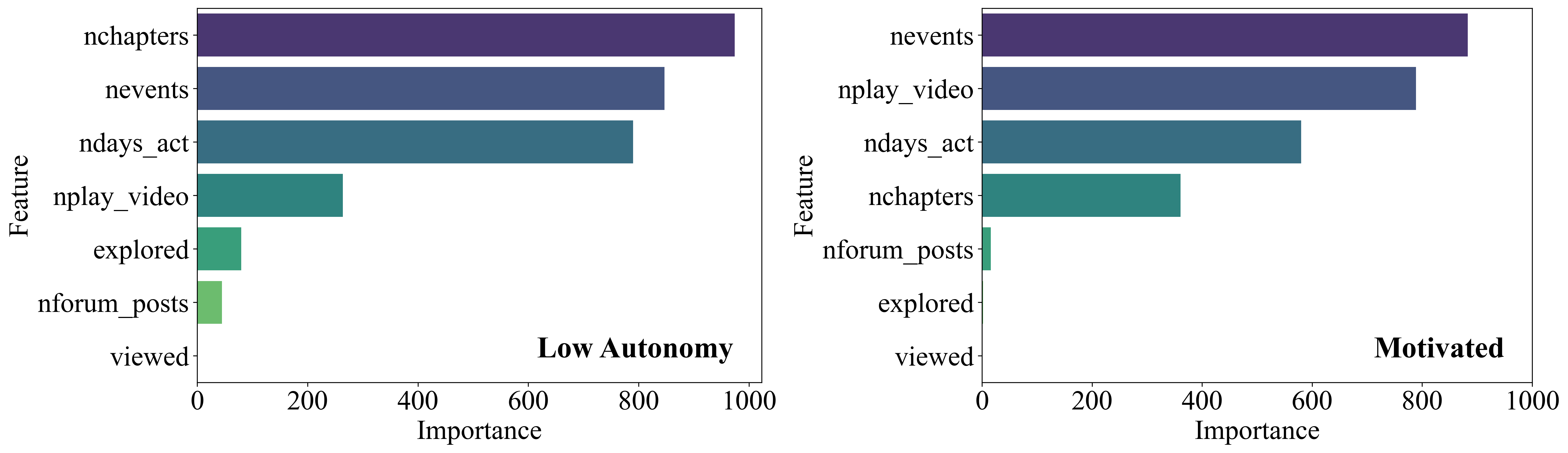}
    \caption{Important features for the low autonomy students (left) and motivated students (right) from the LightGBM algorithm}
    \label{fig_features}
\end{figure}

Given that the integration framework, which identifies two learning patterns, outperforms the direct approach which does not differentiate leaning patterns, the feature importance was calculated for students exhibiting distinct learning patterns using LightGBM algorithm. Figure \ref{fig_features} shows the importance scores of the features for the low autonomy students (left panel) and the motivated students (right panel), respectively, where features are ranked in decreasing order according to their importance scores, with the most important features at the top of the y-axis and the least important at the bottom. The larger the scores of feature importance, the more important the feature. 

In general, nevents, nplay\_video, ndays\_act, and nchapters are significant features for both low autonomy and motivated students, while nforum\_posts, explored, and viewed appear to have very little impact on both groups. The results suggest that interaction frequency, video consumption, and chapter completion may significantly influence the learning success for either group of students, while engagement in forum discussions, exploration of the chapters, and courseware may not.

Though the top four features remain significant for both groups, the rankings shift within different learning patterns. For low autonomy students, nchapters, nevents, and ndays\_act have much higher importance scores than nplay\_video, while for motivated students, nevents, nplay\_video, and ndays\_act have much higher importance scores than nchapters. Therefore, chapter completion seems to be a more important feature for successful online learning for low-autonomy students, while video consumption appears to be more important for motivated students.

\begin{figure}
    \centering
    \includegraphics[width=0.7\linewidth]{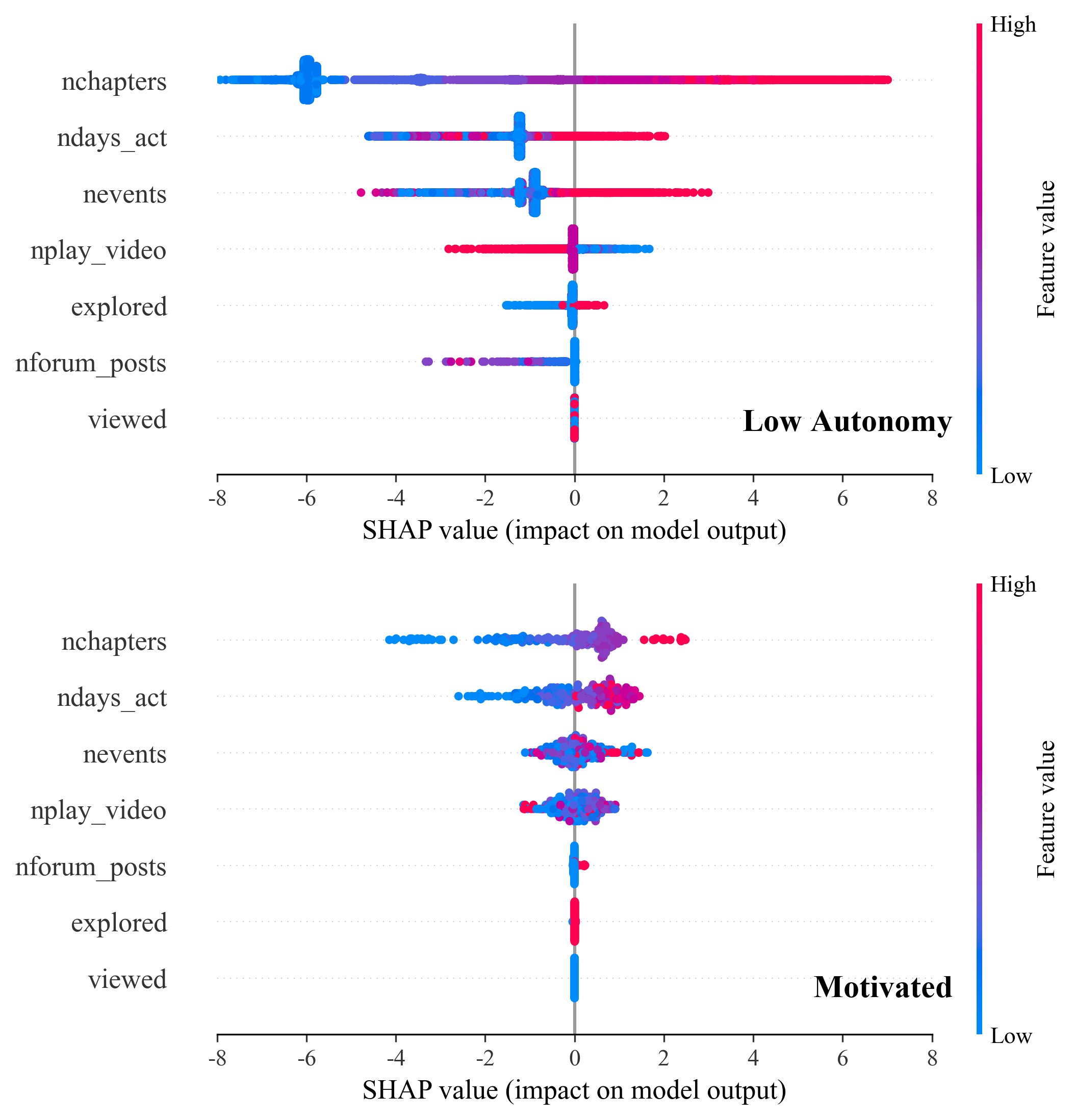}
    \caption{SHAP values of certified for low autonomy students (upper) and motivated students (lower) from the LightGBM algorithm}
    \label{fig_shap}
\end{figure}

Figure \ref{fig_shap} presents the beeswarm plots based on the SHAP values, showing how the top features impact the learning outcomes (successfully obtain a certificate in this emprical study) among low autonomy students (upper panel) and motivated students (lower panel). Each plot showcases one feature per row, with each dot representing the SHAP value for one sample. Positive SHAP values indicate features that increase the prediction, while negative values indicate features that decrease it. Colors indicate the magnitude of feature values, with red representing high values and blue representing low values.

Taking the feature nchapters as an example, which is the most important feature on average, it can be found that low values of nchapters (blue dots) correspond to negative SHAP values (points extending towards the left are increasingly blue), while higher values (red dots) correspond to positive SHAP values (points extending towards the right are increasingly red). The results suggest that students, irrespective of their learning patterns, are more likely to obtain a certificate if they study more chapters. However, the dispersion of nchapters points is greater among low autonomy students compared to motivated students, indicating stronger positive predictions for the former group. Similar trends are found for the features ndays\_act and nevents.

Interesting, for nplay\_video among low autonomy students, higher values of the feature (red dots) are associated with negative SHAP values, whereas lower values with positive SHAP values. In other words, low autonomy students are less likely to obtain a certificate if they watch more videos. Conversely, SHAP values for nplay\_video are densely clustered around zero among motivated students, suggesting that this feature contributes little to predicting obtaining a certificate successfully for these students.

\section{Discussion and Conclusion}

Online learning has become increasingly popular in recent decades, attracting a growing number of learners seeking education through online platforms. Distance education emerged as a critical response to the COVID-19 pandemic, providing a safe and accessible alternative for learning amidst widespread disruptions to traditional educational systems. Alongside this trend, the application of ML algorithms to predict online learning performance has been surging. This study first conducted a scientometric analysis to systematically review studies focusing on predicting learning performance using ML methods over the past years. By leveraging CiteSpace, an excellent tool for bibliometric analysis, we performed keyword co-occurrence analysis to identify research foci and emerging trends within this domain. Our analysis revealed that the most prevalent topics included learning analytics, educational data mining, and ML, indicating a strong emphasis on utilizing advanced methods to predict student performance. 

Further analysis revealed a significant research gap concerning the prediction of learning performance in online learning contexts. Despite the increasing prominence of online education, the number of studies focusing on performance prediction in this domain remains relatively low compared to traditional offline learning contexts. This gap underscores the need for further investigation into online learning performance prediction.

Moreover, while behavior analysis was not identified as a high-frequency term, it has shown promise in enhancing prediction accuracy in several pivotal studies. However, our analysis revealed that only a few studies have integrated behavior analysis with ML algorithms, indicating a significant research gap in this area. Addressing this gap could yield a deeper understanding of students' learning behaviors and their impact on academic outcomes and consequently improve educational practices in both traditional and online learning environments. Therefore, additional research exploring the integration of behavior analysis with ML algorithms is warranted.

This study proposes an innovative integration framework that blends learning behavior analysis with ML algorithms to enhance the prediction of students' online learning performance. Unlike the approach that directly applies ML to the dataset, this framework takes into account the nuanced characteristics of learning behaviors and conducts predictions with respec to different learning patterns of students. Specifically, this framework uses cluster analysis to identify distinct patterns and leverage various ML algorithms to predict performance within each pattern. The studies found that the framework not only enhances the prediction accuracy and precision of ML methods but also provides a deeper understanding of students’ engagement and its associations with learning outcomes.

To demonstrate the effectiveness of the proposed integration framework, this study applies it to an empirical study using a real dataset from the prominent learning platform edX. Within this dataset, two primary learning patterns are identified: low autonomy students and motivated students. The low autonomy students, accounting for the majority proportion of the sample, demonstrate a low certification rate, with a relatively low level of interaction with the course material. In contrast, the motivated students exhibit a significantly higher certification rate, with students displaying substantially higher levels of engagement with the course content. These students access the courseware more frequently, study a greater number of chapters, and actively participate in various course activities such as interactions, video watching, and forum discussions. The stark differences between the two groups of students underscore the importance of considering learning behavior patterns in predicting online learning performance. It was found that there are statistically significant age and gender differences in learning patterns; however, these differences are ignorable due to their small effect size.

The results of applying the integration framework for prediction in the empirical study are promising. For low autonomy students, the framework demonstrates nearly perfect prediction performance. While the prediction performance for motivated students cannot be considered perfect, it is nevertheless satisfactory. The results indicate that the framework can effectively predict the performance of students with varying levels of autonomy and motivation. It was found that the XGBoost method outperforms the other six ML algorithms. In particular, the XGBoost method yields highly satisfactory performance for low-autonomy students, and it also maintains a higher level of performance than other methods for motivated students.

Furthermore, to reveal the advantages of the integration framework, this study conducts a comprehensive comparison between the integration framework and that of directly applying ML methods without learning behavior analysis, using a range of evaluation metrics. The results consistently demonstrate the superiority of the integration framework over the direct approach. On one hand, the integration framework shows significant improvement in prediction accuracy and precision across various metrics, especially for the best-performing XGBoost method. On the other hand, the framework can improve prediction for motivated students, for whom the ML algorithms show relatively lower levels of prediction performance, and for the RF method, which is the worst-performing method. The findings indicate that the integration framework can further enhance the predictive performance of ML methods, leading to even more accurate and reliable predictions of students' online learning performance.

This study analyzes the importance of various learning behaviors using the LightGBM algorithm with the SHAP values for students exhibiting different learning patterns. The results identify that interaction frequency, video consumption, and chapter completion significantly influence the likelihood of successfully obtaining a certificate, whereas engagement in forum discussions and exploration of course material may not. In addition, chapter completion is the most important and positive predictor for both low autonomy and motivated students' success, though the impacts of the feature on the prediction is stronger for the low autonomy students than it is for the motivated students. Furthermore, the video consumption seems to be a negative predictor for the low autonomy students. 

The findings have implications for future studies and the practice of online learning. They offer evidences of the effectiveness and the advantages of the integration framework. The benefits extend beyond the already high-performing XGBoost method to cases where ML methods show relatively lower prediction performance, such as with motivated students, and even to the worst-performing RF method. Therefore, educators and researchers should acknowledge the importance of integrating learning behavior analysis with ML algorithms to enhance predictive performance, and he integration framework proposed in this work represents a promising avenue for future studies. 

The findings also suggest that students' levels of engagement and interaction significantly influence their success in online courses. However, it seems that the efforts dedicated to forum discussion and exploring course materials, particularly among students with low autonomy, may not contribute to completing courses and attaining a certificate successfully. Furthermore, video consumption appears to negatively impact certificate attainment for low autonomy students. In contrast to previous findings \citep[e.g.,][]{AlShabandar2017}, which suggest a strong and positive correlation between clickstream actions and learner outcomes overall, this study implies that this relationship may not hold true for low autonomy students. Moreover, the results suggest that commonly employed methods in distance learning, such as live streaming and recorded lectures, may not suit students lacking autonomy. Particularly when online course videos are poorly designed and made, students may encounter poor learning experiences. Hence, institutions and online course providers are recommended to offer personalized support tailored to individual learner needs.

It is important to acknowledge several limitations of the current study. Firstly, this study utilized the dataset from edX, which may not fully reflect the diversity and complexity within online learning environments. Hence, the generalizability of the findings might be limited. Moreover, the behavior analysis in this study involved seven learning behaviors available in edX. Future studies can include more behavioral features, if applicable, to understand students' learning behaviors more thoroughly. Lastly, while the integration framework exhibits enhanced predictive performance, additional research is needed to assess its effectiveness across varied learning contexts (e.g., both online and offline learning environments).

\vspace{12pt}
\noindent \textbf{\large {Declaration of interest}}

All the authors declare that they have no competing interests.

\bibliographystyle{unsrtnat}


\clearpage
\newpage
\appendix
\setcounter{table}{0} \renewcommand{\thetable}{A.\arabic{table}}
\renewcommand\theHtable{Appendix.\thetable}
\setcounter{figure}{0} \renewcommand{\thefigure}{A.\arabic{figure}}
\renewcommand\theHfigure{Appendix.\thefigure}
\section*{Appendix}

\begin{figure}[!htp]
    \centering
    \includegraphics[width=0.8\linewidth]{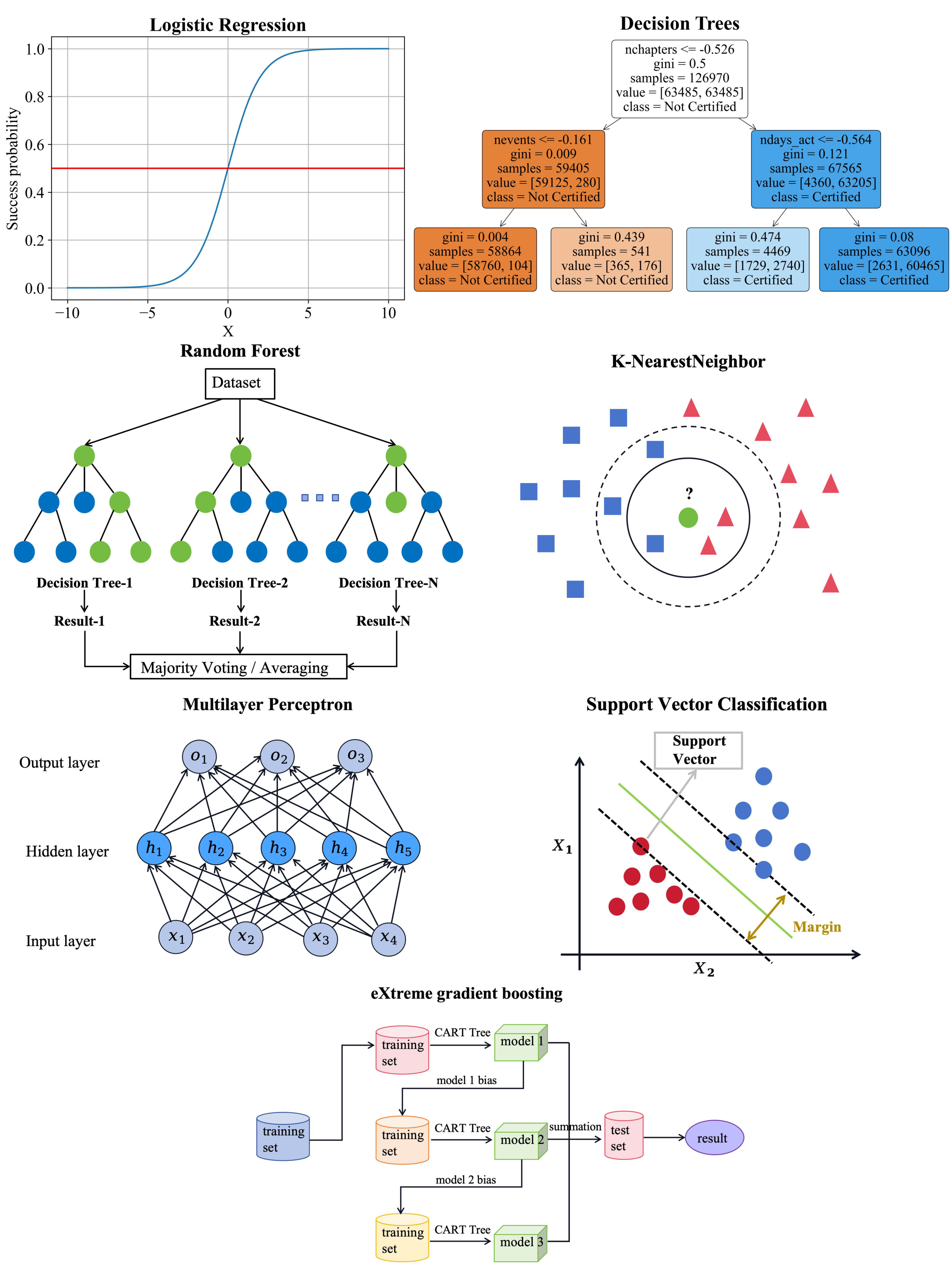}
    \caption{Conceptual illustration of seven machine learning methods}
    \label{fig_ML_conceptual_map}
\end{figure}

\begin{table}[!htp]
\centering
\begin{threeparttable} 
	\caption{Hyperparameterization of machine learning methods for the integration framework and the direct approach}
	\label{tab_hyperpar}	
\begin{tabular}{lllll}
\hline
Method & Hyperparameter & Low autonomy & Motivated & Direct approach\\ \hline
logistic   regression     & regularization factor      & 10    & 0.1   & 10    \\
                          & stopping criterion         & 0.002 & 0.002 & 0.002 \\ \hline
decision tree             & the number of node samples  & 2     & 2     & 2     \\
                          & the minimum number of leaf nodes & 2     & 3     & 1     \\ \hline
random forest             & the number of base learners & 2     & 200   & 2     \\
                          & depth of the tree           & 2     & 7     & 2     \\
                          & the minimum number of leaf nodes & 13    & 12    & 13    \\ \hline
K-Nearest neighbor        & the number of neighbors     & 3     & 20    & 2     \\
                          & the number of leaf nodes    & 2     & 3     & 3     \\ \hline
multilayer perceptron     & activation                  & tanh  & tanh  & tanh  \\
                          & alpha                       & 0.1   & 0.01  & 0.1   \\
                          & hidden layer sizes          & 50    & 50    & 50    \\
                          & kernel                      & rbf   & rbf   & rbf   \\ \hline
support vector classifier & \textit{C}                           & 5     & 5     & 5     \\ \hline
extreme gradient boosting & depth of the tree           & 5     & 7     & 5     \\
                          & the number of base learners & 100   & 60    & 20    \\
                          & maximum number of iterations& 50    & 50    & 50    \\ \hline
\end{tabular}
\begin{tablenotes}
    \small
    \item \textit{Note}. rbf: radial basis function; The columns of low autonomy and motivated are for the integration framework.  
\end{tablenotes}
\end{threeparttable}
\end{table}

\begin{table}[!htp]
\centering
\begin{threeparttable} 
	\caption{Results of clustering under different indices}
	\label{tab_res_clustering}	
\begin{tabular}{lccccccc}
\hline
Index           & 2           & 3          & 4          & 5          & 6          & 7          & 8          \\ \hline
KL$^\star$	&	4.68	&	0.87	&	\textbf{8.39}	&	0.22	&	0.25	&	3.29	&	1.69	\\
CH	&	\textbf{3140.83}	&	2282.78	&	2021.24	&	1620.06	&	1497.25	&	1778.78	&	1738.16	\\
Hartigan	&	1023.41	&	954.17	&	237.34	&	556.18	&	\textbf{1646.14}	&	640.61	&	439.78	\\
CCC	&	0.84	&	7.71	&	24.65	&	17.57	&	26.45	&	86.10	&	\textbf{108.19}		\\
Scott	&	13379.52	&	25288.22	&	44755.92	&	53512.17	&	58746.70	&	\textbf{58746.70}	&	80593.73	\\
Marriot	&	8.42E+43	&	4.28E+43	&	\textbf{6.67E+42}	&	3.49E+42	&	2.61E+42	&	2.09E+41	&	3.02E+41	\\
TrCovw	&	3.78E+07	&	\textbf{2.92E+07}	&	2.71E+07	&	2.36E+07	&	2.21E+07	&	1.79E+07	&	1.46E+07	\\
Tracew	&	66127.13	&	58625.53	&	52376.20	&	50866.40	&	47557.99	&	\textbf{39437.02}	&	36510.83	\\
Friedman	&	17.19	&	20.57	&	33.24	&	\textbf{53.82}	&	53.27	&	63.75	&	59.15	\\
Silhouette	&	\textbf{0.73}	&	0.22	&	0.26	&	0.24	&	0.25	&	0.31	&	0.32	\\
Ratkowsky	&	0.30	&	\textbf{0.31}	&	0.29	&	0.27	&	0.26	&	0.26	&	0.26	\\
Ball	&	33063.57	&	\textbf{19541.84}	&	13094.50	&	10173.28	&	7926.33	&	5633.86	&	4563.85	\\
Ptbiserial	&	\textbf{0.76}	&	0.31	&	0.29	&	0.28	&	0.25	&	0.29	&	0.30	\\
Dunn	&	\textbf{0.02}	&	0.00	&	0.01	&	0.00	&	0.00	&	0.00	&	0.00	\\ \hline
Rubin$^\#$	&	1.39	&	1.57	&	1.76	&	1.81	&	1.94	&	\textbf{2.34}	&	2.52	\\
Cindex	&	0.06	&	0.05	&	0.04	&	0.04	&	0.04	&	0.03	&	\textbf{0.03}	\\
DB	&	\textbf{1.10}	&	1.92	&	1.74	&	2.00	&	1.77	&	1.49	&	1.49	\\
McClain	&	\textbf{0.02}	&	0.63	&	1.22	&	1.52	&	2.10	&	1.99	&	2.11	\\
SDindex	&	6.60	&	5.94	&	4.97	&	4.58	&	4.21	&	4.12	&	\textbf{3.86}	\\
SDbw  &	4.44	&	3.74	&	2.98	&	2.30	&	1.91	&	1.88	&	\textbf{1.73}    \\ \hline
Duda$^\dag$	&	\textbf{1.75}	&	2.11	&	1.40	&	1.91	&	1.12	&	1.70	&	2.05	\\
Pseudot2	&	\textbf{-2028.94}	&	-1404.77	&	-876.24	&	-1222.47	&	-61.55	&	-612.70	&	-947.16	\\
Beale	&	\textbf{-3.50}	&	-4.28	&	-2.31	&	-3.87	&	-0.87	&	-3.36	&	-4.19	\\
Frey	&	\textbf{21.20}	&	0.99	&	0.80	&	2.38	&	-0.11	&	0.11	&	0.26	\\   \hline
\end{tabular}
\begin{tablenotes}
    \small
    \item \textit{Note}. The values in bold indicate the suggested optimal number of clusters. $^\star$: The selection criteria are based on the maximum value of the index or the maximum difference between hierarchy levels of the index; $^\#$: The selection criteria are based on the minimum value of the index or the minimum value of the second difference between levels of the index; $\dag$: The selection criteria are based on a critical value.  
\end{tablenotes}
\end{threeparttable}
\end{table}

\end{document}